\documentclass[aps,prb,groupedaddress, twocolumn]{revtex4}
\usepackage{amsmath}
\usepackage{amssymb}
\usepackage{mathrsfs}
\usepackage{graphicx}
\usepackage{color}
\usepackage{tikz}
\usepackage{pgffor}
\usepackage{verbatim}
\usepackage{pstricks}
\usepackage{pst-3dplot}
\usepackage[colorlinks, breaklinks=true,linkcolor=cyan, citecolor=cyan, linktocpage=true]{hyperref}
\renewcommand{\>}{\rangle}
\newcommand{\bpm}{\begin{pmatrix}}
\newcommand{\epm}{\end{pmatrix}}

\begin{document}

\title{Topological invariants for gauge theories and symmetry-protected topological phases}

\author{Chenjie Wang}
\affiliation{James Franck Institute and Department of Physics, University of Chicago, Chicago, Illinois 60637, USA}

\author{Michael Levin}
\affiliation{James Franck Institute and Department of Physics, University of Chicago, Chicago, Illinois 60637, USA}

\date{\today}

\begin{abstract}
We study the braiding statistics of particle-like and loop-like excitations in 2D and 3D gauge theories with finite, Abelian gauge group. The gauge theories that we consider are obtained by gauging the symmetry of gapped, short-range entangled, lattice boson models. We define a set of quantities --- called {\it topological invariants} --- that summarize some of the most important parts of the braiding statistics data for these systems. Conveniently, these invariants are always Abelian phases, even if the gauge theory supports excitations with non-Abelian statistics. We compute these invariants for gauge theories obtained from the exactly soluble group cohomology models of Chen, Gu, Liu and Wen, and we derive two results. First, we find that the invariants take different values for every group cohomology model with finite, Abelian symmetry group. Second, we find that these models exhaust all possible values for the invariants in the 2D case, and we give some evidence for this in the 3D case. The first result implies that every one of these models belongs to a distinct SPT phase, while the second result suggests that these models may realize all SPT phases. These results support the group cohomology classification conjecture for SPT phases in the case where the symmetry group is finite, Abelian, and unitary.
\end{abstract}
\maketitle

%\makeatletter
%\def\l@subsection#1#2{}
%\def\l@subsubsection#1#2{}
%\makeatother
%\setcounter{tocdepth}{0}

%\tableofcontents

\section{Introduction}

Topological insulators\cite{hasan10, qi11} are a special case of symmetry protected topological (SPT) phases\cite{gu09, pollmann10,fid11,chen11,chen11b,schuch11,chen12,chen13}. These phases can occur in quantum many-body systems of arbitrary dimension and arbitrary symmetry. By definition, a gapped quantum many-body system belongs to a (nontrivial) SPT phase if it satisfies three properties. First, the Hamiltonian is invariant under some set of internal symmetries, none of which are broken spontaneously. Second, the ground state is short-range entangled: that is, it can be transformed into a product state or atomic insulator using a local unitary transformation\cite{verstraete05,vidal07,chen10}. Third, the ground state cannot be continuously connected with a product state, by varying some parameter in the Hamiltonian, without breaking one of the symmetries or closing the energy gap. In addition, nontrivial SPT phases typically have robust boundary modes\cite{hasan10, qi11,chen11c,levin12,lu12,vishwanath13} analogous to that of topological insulators, but this property is not part of the formal definition.

Chen, Gu, Liu, and Wen\cite{chen13} have proposed a general classification scheme for SPT phases built out of bosons. Their classification scheme is based on their construction of a collection of exactly soluble lattice boson models of arbitrary symmetry and spatial dimension. The authors conjecture that these models --- called group cohomology models --- have two basic properties: (i) every group cohomology model belongs to a distinct SPT phase and (ii) every SPT phase can be realized by a group cohomology model. If both properties hold, then it follows logically that there is a one-to-one correspondence between the group cohomology models and SPT phases. In Ref.~\onlinecite{chen13}, the authors assumed this to be the case, and thereby derived a classification scheme for SPT phases based on group cohomology.

While the results and arguments of Ref.~\onlinecite{chen13} represent a major advance in our understanding of SPT phases, they leave several questions unanswered. First, it is not obvious that properties (i)-(ii) hold in general (in fact, property (ii) is known to fail for SPT phases with anti-unitary symmetries\cite{vishwanath13,kapustin14, wen14}). Second, even if these properties do hold at some level, the resulting classification scheme is not completely satisfying since it doesn't tell us how to determine to which SPT phase a microscopic Hamiltonian belongs.

Motivated by these problems, several proposals have been made for how to physically characterize and distinguish SPT phases\cite{lu12,vishwanath13,levin12,threeloop,sule13, cheng13, wen1, wen2, ran14,jwang14,xu13,bi13, ye14,jwang15,barkeshli14,zaletel}. Here, we will focus on the suggestion of Refs.\onlinecite{levin12,threeloop} which applies to 2D and 3D SPT models with \emph{unitary} symmetries. Ref.~\onlinecite{levin12} showed, via a simple example, that one can probe 2D SPT models by gauging their symmetries and studying the braiding statistics of the excitations in the resulting gauge theory. This braiding statistics data is useful because it is invariant under arbitrary symmetry preserving deformations of the Hamiltonian, as long as the energy gap remains open. Therefore, if two SPT models give rise to different braiding statistics, then they must belong to distinct SPT phases.\cite{footnote1} The braiding statistics approach can also be applied to 3D SPT phases.\cite{threeloop,ran14} In that case, after gauging the symmetry, one studies the braiding statistics of the vortex loop excitations in the resulting gauge theory. More specifically, different SPT phases can be distinguished by examining their \emph{three-loop} braiding statistics --- the statistics associated with braiding a loop $\alpha$ around another loop $\beta$, while they are both linked to a third loop $\gamma$.

When considered together, the braiding statistics approach and the group cohomology construction raise several questions:
\begin{enumerate}
\item{Does every group cohomology model lead to distinct braiding statistics?\cite{footnote10}}
\item{Do the group cohomology models exhaust all possible types of braiding statistics that can occur in SPT systems?}
\item{If two SPT models give rise to the same braiding statistics, do they always belong to the same phase?}
\end{enumerate}
The answers to these questions have powerful implications, especially if we can answer them  affirmatively. For example, if we can answer the first question in the affirmative, we can immediately conclude that every group cohomology model belongs to a distinct phase. Likewise, if we can answer the second and third questions in the affirmative, then we can conclude that the group cohomology models realize all possible SPT phases. If we can answer all three questions affirmatively, then it follows that (1) the group cohomology classification is correct and (2) the braiding statistics data provides a universal probe for characterizing and distinguishing different SPT phases with unitary symmetries.

In this work, we consider the first and second questions for the case of 2D and 3D SPT phases with finite Abelian  unitary symmetry group $G=\prod_{i=1}^K \mathbb Z_{N_i}$. We answer the first question in the affirmative and we find evidence that the same is true for the second question, as we explain below.

We obtain our results by focusing on a subset of the braiding statistics data that summarizes some of its most important features (in fact for systems with Abelian statistics, this subset is equivalent to the full set of braiding data, see section \ref{sec:relation}). In the 2D case, this subset consists of $3$ tensors, $\{\Theta_i, \Theta_{ij}, \Theta_{ijk}\}$ that take values in $[0,2\pi]$, where $1\le i,j,k\le K$. In the 3D case, it consists of $3$ tensors $\{\Theta_{i,l}, \Theta_{ij,l}, \Theta_{ijk,l}\}$  with $1\le i,j,k,l\le K$. These tensors --- which we call \emph{topological invariants} --- are defined by considering the Berry phase associated with certain composite braiding processes of vortices or vortex loops. Conveniently, these Berry phases are always Abelian phases regardless of whether the full set of braiding statistics is Abelian or non-Abelian.

%\begin{figure}[b]
%\includegraphics{fig_mapping1.eps}
%\caption{Schematic relations between the set of SPT models $\mathcal {SPT}$, the subset of group cohomology models
%$\mathcal H\subset \mathcal{SPT}$ and the set of topological invariants $\it \Theta$. The subset $\mathcal H\subset \mathcal{SPT}$ denotes the group cohomology models, and $f$ is a map from $\mathcal{SPT}$ to $\it \Theta$.  }\label{fig_mapping}
%\end{figure}

We report two main results. First, we show that the topological invariants take different values in every group cohomology model. Second, we show that the group cohomology models exhaust all possible values for the topological invariants in the 2D case and we give some evidence for this in the 3D case. Our first result implies that the group cohomology models all belong to distinct phases. Our second result can be interpreted as evidence that the group cohomology models realize all possible SPT phases with finite Abelian unitary symmetry group.

%Fig.~\ref{fig_mapping} schematically illustrates the relation between SPT models and topological invariants. For a given finite Abelian symmetry group $G$ and a spatial dimension $d=2$ or $3$, $\mathcal {SPT}$ denotes the set of all SPT models, $\mathcal H$ denotes the set of group cohomology models, and $\it \Theta$ denotes the set of values that the invariants take in all gauged SPT models. Our results show that there exists a well-defined map $f: \mathcal {SPT} \rightarrow {\it \Theta}$ and that the map $f$ is injective when restricted to $\mathcal H$. Also they show that the map $f$ is surjective when restricted to $\mathcal H$ in the 2D case, and we give some partial evidence for this in the 3D case.

Some of our results have appeared previously in the literature, though in a slightly different form. In particular, in the 2D case, Ref.~\onlinecite{zaletel} introduced invariants similar to ours and showed that the invariants can distinguish all the 2D group cohomology models. Also, much of our analysis of 3D gauge theories is similar to that of our previous work, Ref.~\onlinecite{threeloop}. However, this paper goes further than Ref.~\onlinecite{threeloop} in three key ways. First, we study both Abelian and non-Abelian loop braiding statistics, while Ref.~\onlinecite{threeloop} only studied Abelian statistics. Second, we consider a general finite Abelian symmetry group $\prod_{i=1}^K \mathbb Z_{N_i}$ while Ref.~\onlinecite{threeloop} only considered groups of the form $(\mathbb Z_{N})^K$. Finally, we make a systematic comparison between the topological invariants and the group cohomology classification, while Ref.~\onlinecite{threeloop} only made this comparison in a few examples.

A note on our terminology: throughout the paper, we will refer to gauged SPT models as simply {\it gauge theories}. Also, we will refer to the gauged group cohomology models as {\it Dijkgraaf-Witten models}\cite{dijkgraaf90}. The Dijkgraaf-Witten models were studied long before the discovery of SPT phases, however it can be shown that they are equivalent to the gauged group cohomology models (the equivalence is discussed in Appendix \ref{sec:appd_dijkgraafwitten}).

The rest of the paper is organized as follows. In Sec.~\ref{sec:model}, we introduce the models that we will study, both the general gauged SPT models and the more specific Dijkgraaf-Witten models. In Sec.~\ref{sec:defining_theta}, we discuss the general structure of braiding statistics in gauged SPT models and we define the topological invariants. Next, we compute the topological invariants in 2D and 3D Dijkgraaf-Witten models in Sec.~\ref{sec:dijkgraafwitten}. In Sec.~\ref{sec:distinguishingDW}, we show that the topological invariants take different values in every Dijkgraaf-Witten model. In Sec.~\ref{sec:generalconstraint}, we derive general constraints that the topological invariants must satisfy in any gauged SPT model. In Sec.~\ref{sec:beyondcohomology}, we discuss whether the Dijkgraaf-Witten models exhaust all possible values for the invariants. The relation between the topological invariants and the full set of braiding statistics in the case of Abelian statistics is discussed in Sec.~\ref{sec:relation}. Finally, in Sec.~\ref{sec:conclusion}, we conclude and discuss the implications of our results for SPT phases. The Appendices contain several technical details.

\section{Models}

\label{sec:model}

\subsection{Gauge theories}
\label{sec:model1}

The main systems we will study in this paper are 2D and 3D lattice gauge theories with finite Abelian gauge group, $G = \prod_{i=1}^K\mathbb Z_{N_i}$. More specifically, we will study a particular class of gauge theories that are obtained from a two step construction. The first step of the construction is to pick a 2D or 3D lattice boson or spin model with a global $\prod_{i=1}^K\mathbb Z_{N_i}$ symmetry. This boson model can be quite general, with the only restrictions being that (1) it has local interactions, (2) the symmetry is an internal (on-site) symmetry rather than a spatial symmetry, and (3) its ground state is gapped and short-range entangled --- that is, the ground state can be transformed into a product state by a local unitary transformation. Here, by a local unitary transformation, we mean a unitary tranformation $U$ of the form $U = \exp(i H s)$, where $H$ is a local Hermitian operator and $s$ is a finite constant that does not scale with the system size.\cite{verstraete05,vidal07,chen10} (Note that the transformation $U$ need not commute with the symmetry).

The second step of the construction is to gauge the global symmetry of the lattice boson model and couple it to a dynamical lattice gauge field with group $G$. In appendix \ref{sec:appd_gauging}, we give a precise prescription for how to implement this gauging procedure. This prescription mostly follows the usual minimal coupling scheme\cite{kogut79}. However, there is one nonstandard element that is worth mentioning: our procedure is defined so that the gauge coupling constant is \emph{exactly zero}. More precisely, what we mean by this is that the Hamiltonian for the gauged model commutes with the flux operators that measure the gauge flux through each plaquette in the lattice. This property is convenient because it makes the low energy physics of our models well-controlled. In particular, using this property it can be shown that the gauge theories constructed via our gauging procedure are guaranteed to be gapped and deconfined as long as the original boson models are gapped and don't break the symmetry spontaneously (see appendix \ref{sec:appd_gauging}).

The above two step construction defines the class of models that we will study in this paper. From now on, when we use the term {\it gauge theory} we will be referring exclusively to models of this type, unless we state otherwise.

Before concluding this section, we would like to mention that although we find it convenient to use the particular gauging prescription in appendix \ref{sec:appd_gauging}, we don't expect that our results actually depend on the details of the gauging procedure, or on the fact that the resulting gauge theories have zero gauge coupling. Indeed, our results are guaranteed to hold for any model that can be continuously connected one of the above gauge theories without closing the energy gap. We expect that the latter category includes models obtained from generic gauging procedures, as long as the gauge coupling constant is sufficiently small.

\subsection{Dijkgraaf-Witten models}

In part of this work we will study a particular set of exactly soluble gauge theories, known as Dijkgraaf-Witten models\cite{dijkgraaf90} which are obtained by gauging the group cohomology models of Ref. \onlinecite{chen13}. We now briefly review the properties of the group cohomology models and the corresponding Dijkgraaf-Witten models. For the explicit definition of these models, see Appendix \ref{sec:appd_dijkgraafwitten}.

The group cohomology models are exactly soluble lattice boson models that can be defined in any spatial dimension $d$. The basic input needed to construct a $d$-dimensional group cohomology model is a (finite) group $G$ and a $(d+1)$ cocycle $\omega$. Here, an $n$-cocycle $\omega$ is a function $\omega: G^n\rightarrow U(1)$ that satisfies certain conditions. One may define an equivalence relation on cocycles and the equivalence classes are labeled by the elements of the cohomology group $H^{n}[G, U(1)]$ (a brief introduction to group cohomology\cite{brown} is given in Appendix \ref{sec:appd_cohomology}). It can be shown that the models constructed from equivalent cocycles are identical so we will say that the $d$-dimensional group cohomology models are labeled by elements of $H^{d+1}[G, U(1)]$.

Like the group cohomology models, the basic input needed to construct a Dijkgraaf-Witten model in spatial dimension $d$ is a group $G$ and a $(d+1)-$cocycle $\omega$. Also, like the group cohomology models, the Dijkgraaf-Witten models constructed from equivalent cocycles are the same, so we will say that they are labeled by different elements of  $H^{d+1}[G, U(1)]$. Here we will focus on the case $G=\prod_{i=1}^K\mathbb Z_{N_i}$ and $d=2,3$.

%In our later discussions, we will abuse notation and will not distinguish between a cocycle and its equivalence class and hence we will sometimes say that the Dijkgraaf-Witten models are labeled by cocycles.

%In the course of our study below, we will make statements for both the general discrete gauge theories and the specific Dijkgraaf-Witten models. In particular, Sec.~\ref{sec:dijkgraafwitten} is only for Dijkgraaf-Witten models, and other sections in the main text are more or less on general theories.

%Before we move on to braiding statistics, we make a comment on the gauge field. There are two ways to treat the lattice gauge field. In one way, we can treat it as a dynamical field, so that it contains degrees of freedom of the system. In the other way, it can be treated as a static field and can be thought of external parameters. Both treatments are useful. We will mostly work in the first way, however will also refer to the second way in some cases for conceptual clarification.

\subsection{Braiding statistics and phases of SPT models and gauge theories}
\label{sec:model_phase}

Before proceeding further, we briefly review some results on the relationship between SPT models, gauge theories, and braiding statistics. We begin by defining \emph{phases} of SPT models and phases of gauge theories. The former definition is relatively simple: we say that two lattice boson models with the same symmetry group belong to the same SPT phase if they can be continuously connected to one another by varying some parameter in the (symmetry-preserving) Hamiltonian, without closing the energy gap.

Defining phases of gauge theories is more subtle. In fact, there are two inequivalent ways to define this concept, both of which have their merits. In the first definition, two gauge theories belong to the same phase if they can be continuously connected by varying some parameter in the (gauge invariant) Hamiltonian without closing the energy gap. In the second definition, not only do we require the existence of an interpolating Hamiltonian with an energy gap, but we also demand that the interpolating Hamiltonian has \emph{vanishing gauge coupling} ---  that is, the Hamiltonian must commute with the flux operators that measure the gauge flux through each plaquette in the lattice. While the first definition is very natural if one is interested in gauge theories for their own sake, the second definition is more relevant to the study of SPT phases. In this paper, our primary interest is in SPT phases so we will use the second definition.

%Excitations in gauge theories obey interesting braiding statistics.
In parallel to the two ways of defining phases of gauge theories, there are also two ways to define what is means for two gauge theories to have the ``same'' braiding statistics data. In the first definition, two gauge theories have the same braiding statistics data if one can map the excitations of one gauge theory onto the excitations of the other gauge theory such that the corresponding excitations have identical braiding statistics. In the second definition, the corresponding excitations are required both to have the same braiding statistics \emph{and} the same gauge flux. In this paper, we will use the second definition, since it fits more naturally with our definition of phases of gauge theories.
%for our interest of SPT phases, in correspondence to our choice of the second definition of phases of gauge theories.

%We will say that two gauge theories belong to the same phase in the \emph{presence} of flux conservation if they can be continuously
%connected to one another according to the second definition; we will say that they belong to the same phase in the \emph{absence} of
%flux conservation if they can be connected according to the first definition. If two gauge theories belong to the same phase in the
%presence of flux conservation then they necessarily belong to the same phase in the absence of flux conservation, but the converse is
%not always true. In this paper we will be primarily interested in whether two gauge theories belong to the same phase in the presence
%of flux conservation since this is the question that is most relevant to the study of SPT phases. Therefore, in what follows we will
%always define phases of gauge theories in the presence of flux conservation

With these definitions in mind, we can now discuss some results. An important observation is that if two lattice boson models belong to the same SPT phase, then the corresponding gauged models must also belong to the same phase. To see this, note that our gauging prescription (Appendix \ref{sec:appd_gauging}) maps gapped lattice boson models onto gapped zero-coupling gauge theories; hence, any continuous interpolation between two SPT models can be gauged to give an interpolation between the two corresponding gauge theories.

Another important observation is that if two gauge theories belong to the same phase, then they must have the same braiding statistics. One way to see this is to note that braiding statistics data can only take on discrete values and cannot change continuously. (This discreteness property is known as \emph{Ocneanu rigidity} \cite{kitaev06}). Combining the above two observations, we derive a useful corollary: if two lattice boson models belong to the same SPT phase then they must give rise to the same braiding statistics after gauging their symmetries. The converse of this statement may also be true, but it is not obvious.

%One subtlety is that there are actually two inequivalent definitions for when gauge theories have the same braiding
%statistics. These two definitions parallel the two ways of defining phases of gauge theories. In the first definition, two gauge
%theories have the same braiding statistics if one can map the excitations of one gauge theory onto the
%excitations of the other gauge theory such that the corresponding excitations have identical braiding statistics. In the
%second definition, the corresponding excitations are required to both have the same braiding statistics \emph{and} the same gauge flux.
%In this paper, we will use the second definition of equivalence. If two gauge theories have different braiding statistics in this
%sense, then it is easy to see that they must belong to different phases according to the second definition described above.

%These two ways of defining
%equivalence between braiding statistics are closely connected to the two ways of defining phases of gauge theories. If two gauge theories have
%different braiding statistics according to the first definition then they belong to different phases in the absence of
%flux-conservation; on the other hand, if they have different braiding statistics according to the second definition then they belong
%to different phases in the presence of flux conservation. In what follows, we will always assume the second definition of braiding
%statistics unless we state otherwise.

\section{Defining the topological invariants}
\label{sec:defining_theta}

In this section, we construct a set of topological invariants for gauge theories with gauge group $G = \prod_{i=1}^K \mathbb Z_{N_i}$. (Here, when we say ``gauge theory'', we mean a gauge theory of the type discussed in section \ref{sec:model}). These invariants are defined in terms of the braiding statistics\cite{kitaev06,preskill} of the excitations of the gauge theory.  They are denoted by $\Theta_i, \Theta_{ij}, \Theta_{ijk}$ in the 2D case and $\Theta_{i,l}, \Theta_{ij, l}, \Theta_{ijk,l}$ in the 3D case, where the indices $i,j,k,l$ range over $1, \dots, K$.
%A large part of the discussion is devoted to proving that these invariants are well-defined quantities.
For pedagogical purposes, we first define the invariants in the case where the braiding statistics are Abelian, and then discuss the general case (where the statistics may be Abelian or non-Abelian).

\subsection{2D Abelian case}
\label{sec:2dabelian}
We start with the simplest case: we consider 2D gauge theories with group $G = \prod_{i=1}^K \mathbb Z_{N_i}$ and with Abelian braiding statistics.

\subsubsection{Excitations and braiding statistics}
\label{sec:2dabelian1}
%We first discuss the excitation spectrum of these gauge theories. From our knowledge of discrete gauge theories\cite{kogut79}, we know the basic excitations are charges and vortices. The charge excitations can be labeled by integer vectors $q=(q_1, q_2,\dots, q_K)$  where each component $q_i$ takes values in the range $0, 1, \dots, N_i -1$. This label describes the amount of gauge charge carried by the charge excitation. Vortices are excitations that carry gauge flux. Gauge flux can be labeled by a vector $\phi = (\phi_1, \phi_2, \dots, \phi_K)$ where the component $\phi_i$ is a multiple of $\frac{2\pi}{N_i}$ modulo $2\pi$. While the charge excitations are uniquely characterized by $q$, it is important to keep in mind that there are multiple types of vortices carrying the same flux $\phi$: these vortices can be obtained from one another by attaching charge excitations. (Without causing any confusion, we will use $q$ to denote both a charge excitation and its gauge charge. We will use Greek letters $\alpha,\beta,\gamma,\dots$ to denote vortices as well as general excitations throughout the paper.)

We first discuss the excitation spectrum of these gauge theories. In general, every excitation $\alpha$ in a discrete gauge theory\cite{kogut79} can be labeled by the amount of gauge flux $\phi_\alpha$ that it carries. In our case, the gauge flux $\phi_\alpha$ can be described by a $K$-component vector $\phi_\alpha = (\phi_{1\alpha}, \phi_{2\alpha}, \dots, \phi_{K\alpha})$ where each component $\phi_{i\alpha}$ is a multiple of $\frac{2\pi}{N_i}$, and is defined modulo $2\pi$. Excitations can be divided into two groups: \emph{charge} excitations that carry vanishing gauge flux and \emph{vortex} excitations that carry nonzero gauge flux.

As far as their topological properties go, charge excitations are uniquely characterized by their gauge charge $q=(q_1, q_2,\dots, q_K)$ where each component $q_i$ is defined modulo $N_i$. In contrast, vortex excitations are \emph{not} uniquely characterized by the amount of gauge flux that they carry: in fact, there are $|G| = \prod_{i=1}^K N_i$ different types of vortices carrying the same flux $\phi$. All of these vortices can be obtained by attaching charge excitations
%\cite{footnote-charge}
to a fixed reference vortex with flux $\phi$. Throughout this paper, we will use Greek letters $\alpha, \beta, \gamma,\dots$ to denote vortices as well as general excitations, and we will use the letter $q$ to denote both a charge excitation and its gauge charge.

Before proceeding further, we would like to point out a possible source of confusion: given what we have said about the different types of vortices, it is tempting to try to label vortex excitations by \emph{both} their gauge flux and their gauge charge. The problem with this approach is that we do not know any physically meaningful way to define the absolute charge carried by a vortex excitation in a discrete gauge theory. Therefore we will avoid using this notion in this paper. Instead, we will only use the concept of \emph{relative} charge: we will say that two vortices $\alpha, \alpha'$ \emph{differ} by charge $q$ if $\alpha'$ can be obtained by attaching a charge excitation $q$ to $\alpha$.

%(A subtle issue is there is no canonical way to define the absolute charge carried by a vortex: only the relative charge of two vortices has a clear physical meaning. See Sec.~\ref{sec:absolutecharge} for details.)

Let us now consider the braiding statistics of the different excitations. There are three different braiding processes to consider: braiding of two charges, braiding of a charge around a vortex, and braiding of two vortices. The first process is easy to analyze: indeed, it is clear that the charges correspond to local excitations in the ungauged short-range-entangled bosonic state. Therefore the charges are all bosons and have trivial (bosonic) mutual statistics. The braiding between a charge and a vortex is also easy to understand, as it follows from the Aharanov-Bohm law. More specifically, the statistical Berry phase $\theta$ associated with braiding a charge $q$ around a vortex with flux $\phi$ is given by
\begin{equation}
\theta = q\cdot \phi,
\end{equation}
where ``$\cdot$'' is the vector inner product. Note that attaching a charge to the vortex does not change the Aharanov-Bohm law since the charges have trivial mutual statistics with respect to one another.

From the above arguments, we see that the charge-charge and charge-vortex statistics are completely fixed by the gauge group, leaving no room for variation. Therefore, the only braiding process that has potential for distinguishing gauge theories with the same gauge group is vortex-vortex braiding. Motivated by this observation, we will define the topological invariants $\Theta_i, \Theta_{ij}$ in terms of the vortex-vortex braiding statistics.

\subsubsection{The topological invariants}
\label{sec:2dabelian2}
Let $\alpha$ be a vortex carrying a unit flux $\frac{2\pi}{N_i}e_i$, where $e_i = (0, \dots, 1, \dots, 0)$ with a $1$ is the $i$th entry and $0$ everywhere else. Let $\beta$ be a vortex carrying a unit flux $\frac{2\pi}{N_j}e_j$. Here, $i$ and $j$ can take any value in $1, \dots, K$. We define
\begin{align}
\Theta_{ij}  = N^{ij}\theta_{\alpha\beta}, \quad \Theta_{i} = N_i \theta_\alpha, \label{2d_Theta}
\end{align}
where $\theta_{\alpha\beta}$ is the mutual statistics between $\alpha$ and $\beta$, $\theta_\alpha$ is the exchange statistics of $\alpha$, and $N^{ij}$ is the least common multiple of $N_i$ and $N_j$ (More generally, throughout the paper, we use $N^{ij\dots k}$ to denote the least common multiple of $N_i, N_j, \dots, N_k$ and use $N_{ij\dots k}$ to denote the greatest common divisor of $N_i, N_j, \dots, N_k$).

For the quantities $\Theta_{ij}$ and $\Theta_i$ to be well-defined, we need to check that $N^{ij} \theta_{\alpha \beta}$
and $N_i \theta_\alpha$ only depend on $i$ and $j$, and not on the choice of the vortices $\alpha, \beta$. To see that this is the case, imagine that we replace $\alpha, \beta$ with some other vortices $\alpha', \beta'$ carrying flux $\frac{2\pi}{N_i}e_i$ and $\frac{2\pi}{N_j}e_j$. Then clearly the vortices $\alpha$ and $\alpha'$ differ only by the attachment of charge, as do $\beta$ and $\beta'$. Therefore, according to the Aharonov-Bohm law, the change in $\Theta_{ij}$ that occurs when we replace
$\alpha \rightarrow \alpha'$, $\beta \rightarrow \beta'$ is
\begin{align}
\Theta_{ij} \rightarrow \Theta_{ij} + 2\pi N^{ij}\left(\frac{x}{N_i} + \frac{y}{N_j} \right) \label{eqxy}
\end{align}
where $x,y$ are integers that describe the amount of type-$i$ and type-$j$ charge that is attached to $\beta$ and $\alpha$, respectively.
But $N^{ij}$ is divisible by both $N_i$ and $N_j$ so we see that this replacement does not change $\Theta_{ij}$ modulo $2\pi$. Similarly, the Aharonov-Bohm law tells us that the change in $\Theta_i$ that occurs when we replace $\alpha \rightarrow \alpha'$ is
\begin{align}
\Theta_i \rightarrow \Theta_{i} + 2\pi N_i\frac{z}{N_i}, \label{eqz}
\end{align}
where $z$ is the type-$i$ charge that is attached to $\alpha$. Thus $\Theta_i$ is also unchanged modulo $2\pi$. We conclude that the quantities $\Theta_{ij}$ and $\Theta_i$ are both well-defined.

In addition to being well-defined, it is possible to show that $\Theta_{ij}$ and $\Theta_i$ have another nice property: they contain the same information as the full set of braiding statistics. We will derive this result in Sec.~\ref{sec:relation}.

\subsection{2D general case}
\label{sec:2dnonabelian}
In this subsection, we move on to general 2D gauge theories with gauge group $G = \prod_{i=1}^K \mathbb Z_{N_i}$. Unlike the previous section, we do not assume that the braiding statistics of the excitations is Abelian. This additional generality is important because, contrary to naive expectations, gauge theories with Abelian gauge groups can sometimes have excitations with non-Abelian statistics. For example, this phenomenon occurs in 2D $\prod_{i=1}^K \mathbb Z_{N_i}$ Dijkgraaf-Witten models when $K\ge 3$ (c.f. Ref.~\onlinecite{propitius95}).

In the general case, we will define three topological invariants $\Theta_i, \Theta_{ij}, \Theta_{ijk}$. The first two $\Theta_i, \Theta_{ij}$ reduce to those defined in (\ref{2d_Theta}) when restricted to Abelian statistics. The third invariant $\Theta_{ijk}$ is new to the non-Abelian case, and vanishes in the Abelian case.

\subsubsection{General aspects: excitations, fusion rules, and braiding statistics}
\label{sec:2dnonabelianfusion}

Many features of the Abelian case carry over to the general case without change. First, we can still label every excitation $\alpha$ by the amount of gauge flux $\phi_\alpha=(\phi_{1\alpha}, \dots, \phi_{K\alpha})$ that it carries, where $\phi_{i\alpha}$ is a multiple of $\frac{2\pi}{N_i}$ and is defined modulo $2\pi$. Also, we can still divide excitations into two groups: {\it charges}, that carry vanishing flux, and {\it vortices} that carry nonzero flux. Charge excitations are still characterized uniquely by their gauge charge $q=(q_1, \dots, q_K)$ with $q_i$ defined modulo $N_i$, while vortices are still characterized \emph{non-uniquely} by their gauge flux. Finally, charges are still Abelian particles with trivial charge-charge statistics, and with charge-vortex statistics given by the Aharonov-Bohm law: $\theta = q \cdot \phi$ where $\phi$ is the gauge flux carried by the vortex. The main new element in the general case is that vortices can be non-Abelian, i.e., they can have non-Abelian fusion rules and non-Abelian braiding statistics with one another\cite{kitaev06,preskill}.

While the possibility of non-Abelian vortices complicates our analysis, we can still make some general statements about the fusion rules and braiding statistics in these systems. In what follows, we focus on the fusion rules, and we list some properties which will be useful in our later arguments (see Appendix \ref{sec:appd_fusionprop} for proofs and details). To begin, imagine we fuse together two excitations $\alpha$ and $\beta$. In general, there may be a number of possible fusion outcomes corresponding to other excitations $\gamma$:
\begin{equation}
\alpha \times \beta = \sum_\gamma N_{\alpha\beta}^\gamma \gamma ,\label{fusion1}
\end{equation}
where $N_{\alpha\beta}^\gamma $ is the dimension of the fusion space $\mathbb V_{\alpha\beta}^\gamma$. One property of these fusion rules is that
\begin{equation}
\phi_{\gamma} = \phi_{\alpha}+\phi_\beta
\end{equation}
for any fusion product $\gamma$.
In particular, if $\phi_\alpha+\phi_\beta=0$, then all the $\gamma$'s that appear on the right hand side of (\ref{fusion1}) are pure charges.

A second property is that the fusion of a charge $q$ and an excitation $\alpha$ always results in a single excitation
\begin{equation}
\alpha\times q = \alpha' \label{fusion2}
\end{equation}	
where $\alpha'$ is not necessarily distinct from $\alpha$ and $\phi_{\alpha'}=\phi_\alpha$.
A third property is that if two excitations $\alpha,\alpha'$ have the same flux, $\phi_{\alpha'}=\phi_\alpha$, then there exists at least one charge $q$ with
$\alpha' = \alpha \times q$.

To describe the final property, let $\alpha$ and $\beta$ be two excitations, and let $\gamma$ be one of their fusion channels. Let $\alpha'$ and $\beta'$ be two other excitations with $\phi_{\alpha'} = \phi_{\alpha}$ and $\phi_{\beta'} = \phi_\beta$, and let $\gamma'$ be one of their fusion channels. The final property states that there exist charges $q_1$ and $q_2$ such that $\alpha' = \alpha \times q_1$, $\beta' = \beta \times q_2$ and $\gamma' = \gamma \times q_1 \times q_2$.

%More generally, given $n$ vortices $\alpha_1, \dots, \alpha_n$ in a particular fusion state in $\mathbb V_{\alpha_1\dots\alpha_n}^{\gamma}$, one can construct a complete set of states for every other fusion channel $\mathbb V_{\alpha_1'\dots\alpha_n'}^{\gamma'}$ by fusing the vortices with different charges.

\subsubsection{The topological invariants}

Similarly to the Abelian case, we define the topological invariants $\Theta_i, \Theta_{ij}, \Theta_{ijk}$ in terms of the braiding statistics of vortices. Let $\alpha,\beta,\gamma$ be three vortices carrying unit fluxes $\frac{2\pi}{N_i}e_i, \frac{2\pi}{N_j}e_j, \frac{2\pi}{N_k}e_k$ respectively. The topological invariants $\Theta_i, \Theta_{ij}, \Theta_{ijk}$ are defined as follows.\\

\begin{figure}
\includegraphics{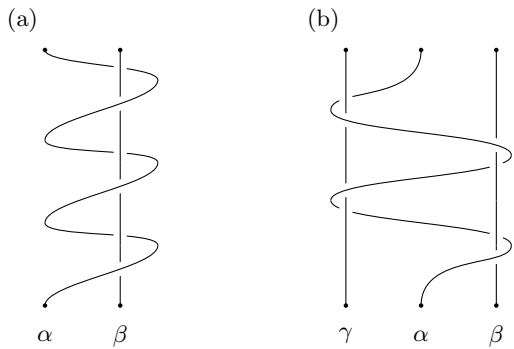}
\caption{Space-time trajectories of the vortices in the braiding processes associated with $\Theta_{ij}$ [panel (a); $N^{ij}=3$] and $\Theta_{ijk}$ [panel (b)]. The arrow of time is upward.}\label{fig_spacetimetraj}
\end{figure}

\noindent{\bf Definitions:}
\begin{itemize}
\item $\Theta_i = 2\pi N_i s_\alpha$, where $s_\alpha$ is the topological spin of $\alpha$;

\item $\Theta_{ij}$ is the Berry phase associated with braiding $\alpha$ around $\beta$ for $N^{ij}$ times;

\item $\Theta_{ijk}$ is the Berry phase associated with the following braiding process: $\alpha$ is first braided around $\beta$, then around $\gamma$, then around $\beta$ in the opposite direction, and finally around $\gamma$ in the opposite direction.
\end{itemize}
Fig.~\ref{fig_spacetimetraj} shows the space-time trajectories of the vortices in the braiding processes of $\Theta_{ij}$ and $\Theta_{ijk}$. We note that the definitions of $\Theta_i$ and $\Theta_{ij}$ reduce to our previous definitions (\ref{2d_Theta}) in the Abelian case, since $2\pi s_\alpha = \theta_\alpha$ for Abelian quasiparticles. We can also see that $\Theta_{ijk} =0$ in the Abelian case.

Before we show that these quantities are well defined, we comment on the definition of $\Theta_i$. In defining $\Theta_i$, we have used the notion of topological spin. The topological spin $s_\alpha$, $0\le s_\alpha<1$, of an anyon $\alpha$ is defined\cite{kitaev06} to be
\begin{equation}
e^{i2\pi s_\alpha} = \frac{1}{d_\alpha} \sum_\gamma d_\gamma  {\rm tr}(R_{\alpha\alpha}^\gamma),
\end{equation}
where $d_\alpha$ and $d_\gamma$ are the quantum dimensions of $\alpha$ and $\gamma$ respectively, $R_{\alpha\alpha}^\gamma$ is the braiding matrix associated with a half braiding (exchange) of two $\alpha$'s, and the summation is over the $\gamma$'s appearing in the fusion product of $\alpha\times\alpha$.

We see that $\Theta_i$ is rather abstract since $s_\alpha$ does not have a direct physical interpretation. In contrast, $\Theta_{ij}, \Theta_{ijk}$ are defined in terms of concrete physical braiding processes. One might wonder if there is a more concrete definition of $\Theta_i$. Indeed, when $N_i$ is even, we find an alternative definition of $\Theta_i$:
\begin{itemize}
\item $\Theta_i$ is the phase associated with exchanging two $\alpha$'s for $N_i$ times, where $\alpha$ is any vortex carrying unit flux $\frac{2\pi}{N_i} e_i$.
%\item Or, it can be defined as the phase associated with braiding $\alpha$ around $\alpha'$ for $N_i/2$ times, where $\phi_\alpha=\phi_{\alpha'}=\phi_i$;
\end{itemize}
This alternative definition provides a direct way to ``measure'' $\Theta_i$ when $N_i$ is even. The equivalence between this alternative definition and the original topological spin definition of $\Theta_i$ follows from two facts: First, exchanging two identical $\alpha$ vortices $N$ times is equivalent to braiding one around the other $\frac{N}{2}$ times. Second, braiding two identical $\alpha$ vortices around one another gives a pure phase $e^{i4\pi s_\alpha}$. (The latter claim, which is less obvious, is proved in Appendix \ref{sec:appd_Raa}).

One problem with the above definition is that it does not make sense when $N_i$ is odd, since in this case the unitary matrix associated with the exchange process is not necessarily a pure phase. Fortunately, we will see later that when $N_i$ is odd, $\Theta_i$ is uniquely determined by $\Theta_{ii}$, and the latter can be directly ``measured'' using a concrete braiding process. This point can be obtained from the constraints on the invariants, which we will study in Sec.~\ref{sec:2dnonabelian_constraint}.

\subsubsection{Proving the invariants are well-defined}
\label{sec:2dnonabelian_proof}

For the invariants to be well defined, we need to prove two points: (i) We need to show that the unitary transformations associated with the above braiding processes are always Abelian phases regardless of the fact that the vortices may be non-Abelian; (ii) We need to show that these Abelian phases are functions of $i,j,k$ only and do not depend on the choice of vortices $\alpha, \beta, \gamma$ as long as they carry fluxes $\frac{2\pi}{N_i}e_i, \frac{2\pi}{N_j}e_j, \frac{2\pi}{N_k}e_k$ respectively. Only point (ii) is needed for showing $\Theta_i$ is well-defined.

Let us start with proving point (ii) for $\Theta_i$. We first review some key properties of topological spin (a detailed discussion of topological spin can be found in Ref.~\onlinecite{kitaev06}.) If $\alpha$ is an Abelian anyon,
%i.e. $d_\alpha=d_\gamma=1$, then there is only one possible fusion outcome $\gamma$, and
$2\pi s_\alpha$ is just the exchange statistics of $\alpha$. In general, $s_{\alpha}=s_{\bar \alpha}$, where $\bar\alpha$ is the anti-particle of $\alpha$. An important property of the topological spin is
\begin{equation}
R_{\beta\alpha}^\gamma R_{\alpha\beta}^{\gamma} = e^{i2\pi(s_\gamma-s_\alpha-s_\beta)}{\rm id}_{\mathbb V_{\alpha\beta}^\gamma}. \label{appdf_braiding}
\end{equation}
Here $\mathbb V_{\alpha\beta}^\gamma$ is the fusion space of $\alpha,\beta$ in the fusion channel $\gamma$ and $R_{\alpha\beta}^{\gamma}$ is the braiding matrix associated with a half braiding of $\alpha$ and $\beta$ in the fusion channel $\gamma$. The notation ${\rm id}_{\mathbb V_{\alpha\beta}^\gamma}$ denotes the identity matrix in the fusion space $\mathbb V_{\alpha\beta}^\gamma$.

With these properties in mind, we now show that $\Theta_i$ is well defined, i.e., we show that
\begin{equation}
2\pi N_i s_{\alpha'}= 2\pi N_i s_\alpha
\label{thetadef}
\end{equation}
for any two vortices $\alpha, \alpha'$ carrying unit flux $\frac{2\pi}{N_i}e_i$. In the first step, we note that we can assume without loss of generality that $\alpha' = q \times \alpha$ for some charge $q$ since according to the properties discussed in Sec.~\ref{sec:2dnonabelianfusion}, any vortex with unit flux $\frac{2\pi}{N_i}e_i$ can be constructed from a fixed vortex by fusing charges with it. To prove the result for this case, we substitute $\beta = q$ and $\gamma = \alpha' = q \times \alpha$ into Eq. (\ref{appdf_braiding}), obtaining
\begin{equation}
R_{q \alpha}^{\alpha'}R_{\alpha q}^{\alpha'}  = e^{i2\pi(s_{\alpha'}-s_\alpha-s_q)} = e^{i2\pi(s_{\alpha'}-s_\alpha)}
\end{equation}
where in the second equality we used the fact that $q$ is a boson so $s_q = 0$. At the same time, we know
\begin{equation}
R_{q \alpha}^{\alpha'}R_{\alpha q}^{\alpha'} = e^{\frac{2\pi}{N_i} \times \text{integer}}
\end{equation}
since the braiding of a charge around a vortex can be computed from the usual Aharonov-Bohm law. Combining these two relations, we
see that $2\pi N_i (s_{\alpha'}-s_\alpha)$ vanishes modulo $2\pi$, proving (\ref{thetadef}).

Next, we prove points (i) and (ii) in the case of $\Theta_{ij}$. Let $\alpha$ and $\beta$ be two vortices carrying unit flux $\frac{2\pi}{N_i}e_i$ and $\frac{2\pi}{N_j}e_j$ respectively. Imagine we perform a full braiding of $\alpha$ around $\beta$ when they are in some fusion channel $\delta$. From the general theory of non-Abelian anyons\cite{kitaev06}, we know that the unitary matrix associated with a full braiding of $\alpha$ around $\beta$ in a fixed fusion channel $\delta$, is a pure phase factor (this result is a corollary of Eq. (\ref{appdf_braiding})). Denoting this phase factor by $e^{i\theta_{\alpha\beta}^\delta}$, the quantity $\Theta_{ij}$ can be computed as
\begin{equation}
\Theta_{ij} = N^{ij} \theta_{\alpha\beta}^\delta
\end{equation}
In order to establish properties (i) and (ii) above, it suffices to show that
\begin{equation}
N^{ij} \theta_{\alpha \beta}^\delta = N^{ij} \theta_{\alpha' \beta'}^{\delta'}
\label{thetaijdef}
\end{equation}
for any other vortices $\alpha',\beta'$ carrying unit flux $\frac{2\pi}{N_i}e_i, \frac{2\pi}{N_j}e_j$, and for any other fusion channel $\delta'$. Indeed, the independence of $N^{ij} \theta_{\alpha \beta}^\delta$ with respect to the fusion channel $\delta$ implies point (i), while the independence of $N^{ij} \theta_{\alpha \beta}^\delta$ with respect to $\alpha, \beta$ implies point (ii).

In fact, it is enough to prove (\ref{thetaijdef}) for the case where $\alpha'=\alpha \times q_1$, $\beta' = \beta \times q_2$, and $\delta' = \delta \times q_1 \times q_2$ for some charges $q_1, q_2$ since according to the general properties discussed in Sec.~\ref{sec:2dnonabelianfusion}, any $\alpha', \beta', \delta'$ can be obtained in this way. But it is easy to prove (\ref{thetaijdef}) in this case. Indeed, from the Aharonov-Bohm law we can deduce the relation
\begin{align}
N^{ij}(\theta_{\alpha' \beta'}^{\delta'} - \theta_{\alpha \beta}^\delta) = 2\pi N^{ij}\left(\frac{q_{2i}}{N_i} + \frac{q_{1j}}{N_j} \right)
\end{align}
where $q_{2i}$ and $q_{1j}$ are integers that describe the amount of type-$i$ and type-$j$ charge carried by $q_2$ and $q_1$, respectively.
We then observe that the expression on the right hand side vanishes modulo $2\pi$ since $N^{ij}$ is divisible by both $N_i$ and $N_j$. This establishes (\ref{thetaijdef}) and proves properties (i) and (ii) for $\Theta_{ij}$.

%To prove that $\Theta_{ijk}$ is an Abelian phase and that it depends on the fluxes $\phi_i, \phi_j, \phi_k$ only, we use the same philosophy as above. By fusing various charges to the vortices, the states in the associated fusion space $\oplus_\delta\mathbb V_{\alpha\beta\gamma}^\delta$ can be exhausted and the all other vortices carrying the fluxes $\phi_i, \phi_j,\phi_k$ can be exhausted.  On the other hand, this process of fusing charges to the vortices should commute with the sequential braiding processes associated with $\Theta_{ijk}$. The commutativity comes after the fact that each individual braiding in that sequence appears twice, one clockwise and the other counterclockwise. With that, everything else follows similarly the proof for $\Theta_{ij}$.

The proof of points (i) and (ii) for $\Theta_{ijk}$ is more technical and is given in Appendix \ref{sec:appd_thetaijk}.

\subsection{3D Abelian case}

\label{sec:3dabelian}

\begin{figure}[b]
\includegraphics{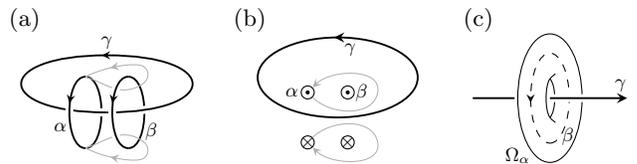}
\caption{Three-loop braiding process. (a) The gray curves show the paths of two points on the moving loop $\alpha$.
(b) Cross-section of the braiding process in the plane that $\gamma$ lies in. (c) A torus $\Omega_\alpha$ is swept out by $\alpha$ during
the braiding, which encloses the loop $\beta$ (dashed circle).}\label{fig_threeloop}
\end{figure}

Having warmed up with the 2D gauge theories, we now consider 3D gauge theories with gauge group $G = \prod_i \mathbb Z_{N_i}$ and with Abelian loop statistics. The discussion that follows is a generalization of the case $G = (\mathbb Z_N)^K$, studied in Ref.~\onlinecite{threeloop}.

\subsubsection{Excitations and three-loop braiding statistics}

%Like the 2D case, the basic excitations in 3D discrete gauge theories are charges and vortices. Again, charges can be labeled by $q=(q_1, \dots, q_K)$ with $0\le q_i < N_i-1$. Vortices carry gauge flux which are again labeled by $\phi=(\phi_1, \dots, \phi_K)$ with the component $\phi_i$ being a multiple of $\frac{2\pi}{N_i}$ modulo $2\pi$. An important difference with respect to the 2D case is that 3D vortices are string-like. In particular, if a string does not end at the surface, it forms a closed loop, which we will refer to a {\it vortex loop} or simply a {\it loop}. Throughout this paper, we assume that the 3D bulk is a closed manifold with no surface, so vortices always form closed loops\cite{footnote2}. As in the 2D case, it is important to keep in mind that there are multiple types of vortices carrying the same flux $\phi$: these vortices can be obtained from one another by attaching charge excitations.

Discrete gauge theories in three dimensions support two types of excitations: charges and vortices. Charge excitations are particle-like and are characterized by the amount of gauge charge $q$ that they carry, where $q=(q_1, \dots, q_K)$ with $q_i$ defined modulo $N_i$. Vortex excitations are string-like and are characterized by the amount of gauge flux $\phi$ that they carry where $\phi =(\phi_{1}, \dots, \phi_{K})$ with the component $\phi_{i}$ being a multiple of $\frac{2\pi}{N_i}$, and defined modulo $2\pi$. We will refer to vortex excitations as {\it vortex loops} or simply {\it loops}, since we will generally assume that the system is defined on a closed manifold with no boundary so that vortex excitations necessarily form closed loops\cite{footnote2}. We will use Greek letters $\alpha,\beta,\gamma$ to denote vortex loop excitations, and will use $\phi_\alpha$ to denote the gauge flux carried by the loop excitation $\alpha$.

As in the 2D case, it is important to keep in mind that while charge excitations are uniquely characterized by their gauge charge, vortex loop excitations are \emph{not} uniquely characterized by their gauge flux: in fact, there are $|G| = \prod_{i=1}^K N_i$ different types of vortex loop excitations carrying the same gauge flux $\phi$. All of these excitations can be obtained by attaching charges
%\cite{footnote-charge}
to a fixed reference loop with flux $\phi$.

Also, just as in 2D, there is some subtlety in defining the absolute charge carried by a vortex loop excitation. Therefore, throughout this paper we will only use the concept of relative charge: we will say that two vortex loops $\alpha$ and $\alpha'$ differ by charge $q$ if $\alpha'$ can be obtained by attaching a charge excitation $q$ to $\alpha$. The only exception to this rule involves \emph{unlinked} vortex loops: when a vortex loop is not linked to any other loops, then there is a natural way to define how much charge it carries: we will say that such a vortex loop is \emph{neutral} if it can be shrunk to a point and annihilated by local operators. Similarly, we will say that an unlinked vortex loop carries charge $q$ if it can be obtained by attaching charge $q$ to a neutral loop.

Let us now consider the braiding statistics of these excitations. There are several types of processes we can consider: braiding of two charges, braiding of a charge around a vortex loop, and braiding involving several vortex loops. As in the 2D case, it is easy to see that the charge-charge statistics are all bosonic and the statistics between a charge $q$ and a vortex loop carrying a flux $\phi$ follow the Aharonov-Bohm law
\begin{equation}
\theta = q\cdot\phi.
\end{equation}
We note that the above Aharonov-Bohm law holds quite generally: it does not depend on the amount of charge attached to the loop, nor on whether the loop is unlinked or linked with other loops.

What is left are braiding processes involving several vortex loops. In general, there are many kinds of loop braiding processes we can consider --- including processes involving two loops\cite{aneziris91,alford92,baez07}, three loops\cite{threeloop,ran14,jwang14,jian14,bi14}, or even more complicated configurations. Here, we will follow Ref.~\onlinecite{threeloop} and focus on the three-loop braiding process
%\cite{footnote3}
depicted in Fig.~\ref{fig_threeloop}, in which a loop $\alpha$ is braided around a loop $\beta$ while both are linked to a third ``base'' loop $\gamma$. Ref.~\onlinecite{threeloop} argued that this three-loop braiding process is a useful probe for characterizing and distinguishing 3D topological phases. As in Ref.~\onlinecite{threeloop}, we denote the statistical phase associated with this braiding process by $\theta_{\alpha\beta,c}$, where $c$ is an integer vector that characterizes the amount of flux carried by $\gamma$. More specifically, $c$ is defined by $\phi_\gamma = (\frac{2\pi}{N_1}c_1, \dots, \frac{2\pi}{N_K}c_K)$. We use the notation $\theta_{\alpha\beta,c}$ rather than $\theta_{\alpha\beta,\gamma}$, because the statistical phase is insensitive to the amount of charge attached to $\gamma$ and depends only on its flux $\phi_\gamma$ which is parameterized by $c$. We will also consider an exchange or half-braiding process, in which two identical loops $\alpha$, both linked to the base loop $\gamma$, exchange their positions. We denote the associated three-loop exchange statistics by $\theta_{\alpha,c}$.

\subsubsection{The topological invariants}

We define our topological invariants, $\Theta_{i,l}$ and $\Theta_{ij,l}$, in terms of the the three-loop braiding statistics of vortex loops. Let $\alpha$ and $\beta$ be two vortex loops carrying unit flux
$\frac{2\pi}{N_i}e_i$ and $\frac{2\pi}{N_j}e_j$, where $e_i=(0,\dots,1,\dots,0)$ with the $i$th entry being $1$ and all others being $0$. Suppose both $\alpha$ and $\beta$ are linked to a third vortex loop $\gamma$ carrying unit flux $\frac{2\pi}{N_l}e_l$. We define
\begin{equation}
\Theta_{ij,l} = N^{ij}\theta_{\alpha\beta,e_l}, \ \Theta_{i,l} = N_i\theta_{\alpha,e_l} \label{3d_Theta}
\end{equation}
Using arguments similar to those for $\Theta_i,\Theta_{ij}$ from Sec.~\ref{sec:2dabelian2}, one can show that the quantities $\Theta_{i,l}, \Theta_{ij,l}$ depend only on $i,j,l$ and not on the choice of vortices $\alpha, \beta, \gamma$. Thus, $\Theta_{ij,l}$ and $ \Theta_{i,l}$ are well-defined quantities.

In addition to being well-defined, it is possible to show that $\Theta_{ij,l}$ and $\Theta_{i,l}$ contain all the information about the three-loop braiding statistics in the gauge theory. We will derive this result in Sec.~\ref{sec:relation}.

\subsection{3D general case}
\label{sec:3dnonabelian}
In this section, we move on to general 3D gauge theories with gauge group $G = \prod_{i=1}^K \mathbb Z_{N_i}$. Unlike the last section, we do not assume that the three-loop braiding statistics is Abelian. We will define three topological invariants $\Theta_{i,l}, \Theta_{ij,l}$ and $\Theta_{ijk,l}$. The first two, $\Theta_{i,l}, \Theta_{ij,l}$, reduce to those defined in (\ref{3d_Theta}) when restricted to Abelian statistics. The third invariant $\Theta_{ijk,l}$ is new to the non-Abelian case, and vanishes in the Abelian case.

\begin{figure}
\includegraphics{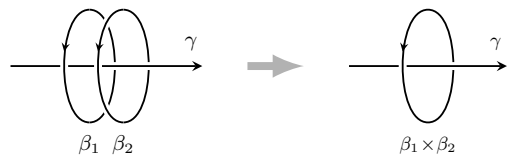}
\caption{Fusion of two loops $\beta_1$ and $\beta_2$, both linked to $\gamma$. We denote this type of fusion by $\beta_1\times\beta_2$. (This is different notation from Ref.~\onlinecite{threeloop}, where this type of fusion was denoted by $\beta_1 + \beta_2$).} \label{fig_fusion}
\end{figure}

\subsubsection{General aspects of non-Abelian loop braiding}
In order to analyze the general case, it is important to recognize the analogy between 3D loop braiding and 2D particle braiding. This analogy can be seen most easily by examining a 2D cross-section of a loop braiding process, as shown in Fig. \ref{fig_threeloop}(b). Here we see that a braiding process involving two loops $\alpha, \beta$ that are linked to a base loop $\gamma$, can be mapped onto a braiding process involving two point-like particles in two dimensions. More generally, any braiding process involving loops $\alpha_1,...,\alpha_N$ that are linked to a base loop $\gamma$ can be mapped onto a braiding process involving $N$ point-like particles in two dimensions. It can be shown that this mapping between 3D loop braiding and 2D particle braiding is one-to-one, so that the 3D braid group for loops (when linked to a base loop) is identical to the 2D braid group for particles.\cite{footnote4} In addition to braiding, there is also a close analogy between fusion processes in two and three dimensions. Just as two particles can be fused together to form another particle, two loops $\alpha, \beta$ that are linked to the same loop $\gamma$ can be fused to form a new loop that is also linked to $\gamma$ (Fig.~\ref{fig_fusion}).

This correspondence between the 2D and 3D cases implies that the algebraic structure of fusion and braiding in 2D anyon theories\cite{kitaev06} can be carried over without change to the theory of 3D loop excitations. In particular, for any loops $\alpha$ and $\beta$ that are both linked with $\gamma$, we can define an associated fusion space $\mathbb V_{\alpha\beta,c}^\delta$, where $\delta$ denotes their fusion channel. Also, we can define an $F$-symbol $F_{\alpha\beta\mu,c}^\delta$ that describes a unitary mapping between two different ways of parameterizing the fusion of three loops
\begin{equation}
F_{\alpha\beta\mu,c}^\delta: \bigoplus_\xi V_{\alpha\beta,c}^\xi\otimes V_{\xi\mu, c}^\delta \rightarrow   \bigoplus_\eta V_{\alpha\eta,c}^\delta \otimes V_{\beta\mu,c}^\eta,
\end{equation}
and that satisfies the pentagon equation\cite{kitaev06}.
Likewise, we can define an $R$-symbol $R_{\alpha\beta,c}^\delta$ which is a unitary transformation
\begin{equation}
R_{\alpha\beta,c}^\delta: V_{\alpha\beta,c}^\delta \rightarrow V_{\beta\alpha,c}^\delta
\end{equation}
and that satisfies the hexagon equation.\cite{kitaev06}
As in the 2D case, the $R$-symbol describes a half-braiding of loops: a full braiding of two loops $\alpha, \beta$ that are in a fusion channel $\delta$ is given by $R_{\beta\alpha,c}^\delta R_{\alpha\beta,c}^\delta$. Finally, we can define quantum dimensions and topological spins of loop excitations. The topological spin of a loop $\alpha$ that is linked to $\gamma$ is given by
\begin{equation}
e^{i2\pi s_{\alpha,c}} = \frac{1}{d_{\alpha,c}}\sum_\delta d_{\delta, c} {\rm tr}(R_{\alpha\alpha,c}^\delta)
\end{equation}
where $d_{\alpha,c}$ and $d_{\delta,c}$ are quantum dimensions.

For all of the above quantities, $\mathbb V_{\alpha\beta,c}^\delta, F_{\alpha\beta\mu,c}^\delta$, etc., the dependence on the base loop $\gamma$ enters through the index `$c$' where $c$ is an integer vector defined by $\phi_\gamma = (\frac{2\pi}{N_1}c_1, \dots, \frac{2\pi}{N_K}c_K)$. The reason we use the notation $\mathbb V_{\alpha\beta,c}^\delta$, etc. rather than $\mathbb V_{\alpha\beta,\gamma}^\delta$, etc. is because it is clear that these quantities depend only on the flux carried by $\gamma$ which is parameterized by $c$.

%For linked loops, there are at least two possible types of fusion. First, two loops $\alpha,\beta$ that are linked to the same loop $\gamma$ can be fused to form a new loop that is also linked to $\gamma$(Fig.~\ref{fig_fusion}a). Second, two loops $\alpha_1,\alpha_2$ that carry the same flux $\phi_{\alpha_1}=\phi_{\alpha_2}$ and are linked to different loops $\gamma_1$ and $\gamma_2$ can be fused to form a new loop that is linked to both $\gamma_1$ and $\gamma_2$ (Fig.~\ref{fig_fusion}b).

%Third, if a loop $\alpha$ is linked to a loop $\gamma$, we can shrink $\alpha$ to a point and fuse it onto $\gamma$.

%The second type of fusion is much less known. It makes a connection between three-loop braiding statistics on different base loops and constrains the possible choices of $F$- and $R$-symbols. We only understand this type of fusion in the case of Abelian statistics. Fortunately, the topological invariants are Abelian phases, so we are able to obtain some conclusions from this type of fusion when discuss the constraints of the topological invariants in Sec.~\ref{sec:3dabelian} and Sec.~\ref{sec:2dnonabelian_constraint}. We leave the detailed properties of this type for future study.

\subsubsection{The topological invariants}
Having established the analogy between 3D loop braiding and 2D particle braiding, we now define the 3D topological invariants using the same approach as in the 2D case. Let $\alpha,\beta,\gamma$ be three vortex loops that are linked with another loop $\sigma$. Suppose that
$\alpha, \beta, \gamma, \sigma$ carry unit flux $\frac{2\pi}{N_i} e_i,\frac{2\pi}{N_j} e_j,\frac{2\pi}{N_k} e_k,\frac{2\pi}{N_l} e_l$, respectively. The topological invariants $\Theta_{i,l}, \Theta_{ij,l}, \Theta_{ijk,l}$ are defined as follows.

%In the case of non-Abelian loop braiding, excitations are again divided into charges and vortex loops. Charges are Abelian particles meaning that they are bosons and their statistics with respect to any vortex loops satisfy the Abelian Aharanov-Bohm law. Vortex loops carry gauge fluxes. They may become non-Abelian, in the sense that the three-loop braiding statistics is non-Abelian. In the absence of the base loop, i.e., the three-loop process reduces to a two-loop process, one can show that the loop braiding must be Abelian (see xxx).

%Let now give the definitions of the topological invariants.  \\

\noindent{\bf Definitions:}
\begin{itemize}
\item $\Theta_{i,l} = 2\pi N_i s_{\alpha,e_l}$, where $s_{\alpha,e_l}$ is the topological spin of $\alpha$ when it is linked to $\sigma$;
\item $\Theta_{ij,l}$ is the Berry phase associated with braiding the loop $\alpha$ around $\beta$ for $N^{ij}$ times, while both are linked to $\sigma$;
\item $\Theta_{ijk,l}$ is the phase associated with the following braiding process: $\alpha$ is first braided around $\beta$, then around $\gamma$, then around $\beta$ in a opposite direction, and finally around $\gamma$ in a opposite direction. Here $\alpha,\beta,\gamma$ are all linked with $\sigma$.
\end{itemize}
Similarly to the 2D case, there is an alternative and more concrete definition of $\Theta_{i,l}$ when $N_i$ is even: $\Theta_{i,l}$ can be defined as the phase associated with exchanging two $\alpha$ loops for $N_i$ times.

We need to prove two points to show these quantities are well-defined: (i) We need to show that the unitary transformations associated with the above braiding processes are always Abelian phases regardless of the fact that the vortex loops may be non-Abelian; (ii) We need to show that these Abelian phases are functions of $i,j,k,l$ only and do not depend on the choice of vortex loops $\alpha, \beta, \gamma, \sigma$ as long as they carry fluxes $\frac{2\pi}{N_i} e_i,\frac{2\pi}{N_j} e_j,\frac{2\pi}{N_k} e_k,\frac{2\pi}{N_l} e_l$. These two properties can be established using similar arguments to those given in the 2D case in Sec.~\ref{sec:2dnonabelian}.

\subsection{Examples}

To see some examples of these invariants, we consider the gauged group cohomology models of Ref.~\onlinecite{chen13} or equivalently, the Dijkgraaf-Witten models of Ref. \onlinecite{dijkgraaf90}. We will compute the invariants for these models in the next two sections. All the results listed below follow from two formulas which we will derive later, namely
(\ref{theta_value_2d1}-\ref{theta_value_2d3}) and (\ref{theta_value_3d1}-\ref{theta_value_3d3}).

The simplest nontrivial example is given by the $2D$ Dijkgraaf-Witten models with symmetry group $G = \mathbb Z_2$. In this case, $H^3(\mathbb Z_2, U(1)) = \mathbb Z_2$ so we can construct two Dijkgraaf-Witten models\cite{levin12}. The only independent invariant in this case is $\Theta_1$, which describes the phase associated with exchanging two identical $\pi$ vortices {\it twice}. The values of $\Theta_1$ in the two Dijkgraaf-Witten models are
\begin{equation}
\begin{array}{ll}
\text{Trivial model:} \quad & \Theta_1 = 0 \nonumber\\
\text{Non-trivial model}: \quad & \Theta_1 = \pi \nonumber
\end{array}
\end{equation}
Importantly, we can see that $\Theta_1$ takes different values in the two models, which proves that they belong to distinct phases.

More generally, for $G = \mathbb Z_N$, we have $H^3(\mathbb Z_N, U(1)) = \mathbb Z_N$, so we can construct $N$ Dijkgraaf-Witten models in this case. Similarly to the $\mathbb Z_2$ case, these models can be distinguished from one another by the topological invariant $\Theta_1$, which takes a different value in $0, \frac{2\pi}{N}, \dots, \frac{2\pi}{N}(N-1)$ for each of the $N$ models.

Another interesting example is given by the 2D Dijkgraaf-Witten models with symmetry group
$G =\mathbb Z_N\times\mathbb Z_N\times \mathbb Z_N$. In this case,
$H^3(\mathbb Z_N \times \mathbb Z_N \times \mathbb Z_N, U(1)) = \mathbb Z_N^7$ so we can construct $N^7$ models.
Interestingly, there are also seven independent topological invariants in this case:
\begin{equation}
\Theta_1, \  \Theta_2, \ \Theta_3, \ \Theta_{12},\  \Theta_{13},\  \Theta_{23}, \ \Theta_{123}. \nonumber
\end{equation}
The invariant $\Theta_1$ is the topological spin of a vortex that carries $\frac{2\pi}{N}(1,0,0)$ flux, multipled by $N$. The invariant $\Theta_{12}$ is the phase associated with braiding a vortex carrying $\frac{2\pi}{N}(1,0,0)$ flux around a
vortex carrying $\frac{2\pi}{N}(0,1,0)$ flux for $N$ times. The invariant $\Theta_{123}$ is the phase associated
with braiding $\frac{2\pi}{N}(1,0,0)$ flux around $\frac{2\pi}{N}(0,1,0)$ flux in the counterclockwise direction,
then around $\frac{2\pi}{N}(0,0,1)$ flux in the counterclockwise direction, then around the same
$\frac{2\pi}{N}(0,1,0)$ flux and $\frac{2\pi}{N}(0,0,1)$ flux in the clockwise direction. The meanings of the other
invariants are similar. The invariant $\Theta_{123}$ is an indicator of non-Abelian statistics: if $\Theta_{123}=0$, the
corresponding statistics is Abelian; otherwise, the statistics is non-Abelian. We find that all seven invariants take values in
\begin{equation}
0, \ \frac{2\pi}{N}, \ \frac{4\pi}{N} \ \dots, \ \frac{2\pi(N-1)}{N} .\nonumber
\end{equation}
and that these values distinguish all of the $N^7$ 2D $\mathbb Z_N\times\mathbb Z_N\times \mathbb Z_N$ models.
Again, this result proves that the $N^7$ models each belong to a different phase according to the definition given in Sec.~\ref{sec:model_phase}.

Finally, we consider 3D Dijkgraaf-Witten models with symmetry group $G = \mathbb Z_2\times \mathbb Z_2$. In this case,
$H^3(\mathbb Z_2 \times \mathbb Z_2, U(1)) = \mathbb Z_2^2$ so we can construct $4$ models\cite{chen13b}. We find there are two
independent topological invariants in this case, namely, $\Theta_{1,2}$ and $\Theta_{2,1}$. While there exist other
invariants such as $\Theta_{12,1}, \Theta_{12,2}$, they are not independent, as we show in Sec.~\ref{sec:generalconstraint}. The
invariant $\Theta_{1, 2}$ is the phase associated with exchanging two identical loops that carry a $(\pi, 0)$ flux
while both loops are linked to a third loop that carries a $(0, \pi)$ flux. The meaning of $\Theta_{2,1}$ is similar.
We find that
\begin{align}
\Theta_{1,2} &= 0 \text{ or } \pi, \nonumber\\
\Theta_{2,1} & = 0 \text{ or } \pi. \nonumber
\end{align}
Each of the four combinations of the values occurs in a different model. Hence, once again, the invariants distinguish all of
the 3D $\mathbb Z_2\times \mathbb Z_2$ Dijkgraaf-Witten models and prove that they belong to distinct phases according to the definition given in Sec.~\ref{sec:model_phase}.

\section{The invariants in Dijkgraaf-Witten models}
\label{sec:dijkgraafwitten}
In this section, we compute the topological invariants for all 2D and 3D Dijkgraaf-Witten models with Abelian gauge group
$G = \prod_{i=1}^K \mathbb Z_{N_i}$. We obtain explicit expressions of the invariants in terms of the cocycle $\omega$ that is used to define the Dijkgraaf-Witten model.

\subsection{ 2D topological invariants}

\subsubsection{Review of braiding statistics in 2D Dijkgraaf-Witten models}

In this section, we summarize some previously known results on the braiding statistics in 2D Dijkgraaf-Witten models. Although these results do not provide an {\it explicit} formula for the braiding statistics in Dijkgraaf-Witten models, they do the next best thing: they give a well defined mathematical procedure for how to compute these statistics in terms of the 3-cocycle $\omega$ that defines the model. This procedure involves a mathematical structure known as the {\it twisted quantum double} algebra \cite{dijkgraaf90b}. The twisted quantum double formalism is quite general and can be applied to any finite group $G$, including non-Abelian groups. However, in the following discussion, we will specialize to the case of Abelian $G$, and we will only give a minimal review of the ingredients that are necessary for the computation of the invariants $\Theta_i, \Theta_{ij}, \Theta_{ijk}$. For more details, readers may consult Ref.~\onlinecite{dijkgraaf90b, propitius95}.

The first component of the twisted quantum double formalism is a scheme for labeling quasiparticle excitations.
Let us consider a Dijkgraaf-Witten model corresponding to an Abelian group $G = \prod_{i=1}^K \mathbb Z_{N_i}$ and 3-cocycle $\omega$.
According to the formalism, each excitation in this model can be uniquely labeled by a doublet
\begin{equation}
\alpha = (a, \rho)
\end{equation}
where $a =(a_1,...,a_K)$ is a group element of $G$ with $0\le a_i \le N_i-1$, and $\rho$ is an irreducible projective representation of $G$ satisfying
\begin{equation}
\rho(b)\rho(c) = \chi_a(b, c)\rho(b+c), \label{projective}
\end{equation}
for all $b,c\in G$. Here, $\chi_a$ is a phase factor defined by
\begin{equation}
\chi_a(b,c) = \frac{\omega(a,b,c)\omega(b,c,a)}{\omega(b,a,c)}, \label{slant1}
\end{equation}
and is called the {\it slant product} of $\omega$. The two labels $(a,\rho)$ have a simple physical meaning: the first component $a$ describes the amount of flux
$\phi_\alpha= (\frac{2\pi}{N_1} a_1,...,\frac{2\pi}{N_k} a_k)$ carried by the excitation $\alpha$, while the second component $\rho$ is related to the amount of charge attached to $\alpha$. For a more precise correspondence between the mathematical labels $(a,\rho)$ and the physical notions of gauge flux and gauge charge, we refer the reader to Appendix \ref{appd:correspd}.

The second component of the formalism is a formula for the fusion rules of the excitations. Specifically,
\begin{equation}
(a, \rho) \times (b, \mu) = \sum_\sigma N_{\rho \mu}^\sigma (a+b, \sigma)
\end{equation}
where the fusion multiplicities $N_{\rho \mu}^\sigma$ are computed as follows. First, we define a projective representation $\rho * \mu$ by
\begin{equation}
(\rho * \mu)(g) = \rho(g) \otimes \mu(g) \cdot \chi_{g}(a,b) \label{prodrep}
\end{equation}
for any $g \in G$. The $N_{\rho \mu}^\sigma$ are then defined in terms of the decomposition of $\rho * \mu$ into irreducible projective representations, $\sigma$:
\begin{equation}
\rho * \mu = \bigoplus_\sigma N_{\rho \mu}^\sigma \sigma
\end{equation}

In addition to fusion rules, this formalism provides a convenient way to parameterize the degenerate ground states associated with a collection
of (non-Abelian) excitations. Consider a system of $n$ excitations $\alpha_i = (a_i, \rho_i)$, $i=1,...,n$, and suppose that these excitations fuse to
the vacuum. The ground state manifold associated with these excitations can be obtained in two steps. First, we construct the tensor product
$V = V_1 \otimes \dots \otimes V_n$ where $V_i$ is the vector space on which $\rho_i$ is defined. Then, we project onto the subspace of $V$ that corresponds to the vacuum fusion channel. This projection is implemented by the operator
\begin{equation}
P= \frac{1}{|G|}\sum_{g \in G} \rho_1(g) * \dots * \rho_n(g)
\end{equation}
The degenerate ground states associated with $\alpha_1,...,\alpha_N$ can be parameterized by vectors that lie in the image of $P: V \rightarrow V$.

We are now ready to present a formula for the braiding statistics of the excitations. Suppose that the above system of $n$ excitations are arranged in a line in the order $\alpha_1,...,\alpha_n$. Then, the unitary transformation associated with braiding $\alpha_i$ around its neighbor $\alpha_{i+1}$, is given by\cite{propitius95}
\begin{eqnarray}
B_{\alpha_i \alpha_{i+1}} &=& P \cdot {\rm id}_{V_1} \otimes \dots \otimes {\rm id}_{V_{i-1}} \otimes \rho_i(a_{i+1})\otimes \rho_{i+1}(a_i) \nonumber \\
&\otimes& {\rm id}_{V_{i+2}} \otimes \dots \otimes {\rm id}_{V_{n}} \cdot P, \label{dw_braiding}
\end{eqnarray}
where ${\rm id}_{V_i}$ denotes the identity matrix on the vector space $V_i$ of $\rho_i$. (The exchange statistics for the excitations can be computed in a similar fashion, but we will not discuss them
 here as they are not necessary for our purposes).

The final result we will need is an expression for the topological spin. According to the twisted quantum double formalism, the topological spin of an excitation $\alpha=(a, \rho)$ is given by\cite{propitius95}
\begin{equation}
e^{i2\pi s_\alpha} = \frac{1}{{\rm dim}(\rho)}{\rm tr}\rho(a), \label{dw_exchange}
\end{equation}
where ${\rm dim}(\rho)$ is the dimension of the representation $\rho$. This formula can equivalently be written as
\begin{equation}
e^{i 2\pi s_\alpha} \ {\rm id}_V = \rho(a) \label{dw_exchange2}
\end{equation}
where ${\rm id}_V$ is the identity matrix in the vector space $V$ of $\rho$. The reason that (\ref{dw_exchange}) is equivalent
to (\ref{dw_exchange2}) is that $\rho(a)$ is always a pure phase. Indeed, this property follows from Schur's lemma and the observation that
$\chi_{a}(a,b)=\chi_{a}(b,a)$ so that $\rho(a)$ commutes with any other matrix $\rho(b)$. (As an aside, we note that the fact that $\rho(a)$ is a pure phase is consistent with the results in Appendix \ref{sec:appd_Raa}.)

\subsubsection{Explicit formulas for the invariants}
We now compute the invariants $\Theta_i$, $\Theta_{ij}$ and $\Theta_{ijk}$ for a 2D Dijkgraaf-Witten model with group
$G = \prod_{i=1}^K \mathbb Z_{N_i}$ and 3-cocycle $\omega$. Let $\alpha, \beta, \gamma$ be three vortices carrying unit flux $2\pi e_i/N_i$, $2\pi e_j/N_j$ and $2\pi e_k/N_k$ respectively. Using the notation from the previous section, we can label these vortices as $\alpha=(e_i, \rho)$, $\beta=(e_j, \mu)$ and $\gamma=(e_k, \nu)$
for some projective representations $\rho, \mu, \nu$. We will denote the vector spaces associated with $\rho,\mu,\nu$ by $V,W,X$.
To compute $\Theta_i$, $\Theta_{ij}$ and $\Theta_{ijk}$, we need to find the topological spin of these vortices and to analyze various
braiding processes involving them.

We begin with $\Theta_i$. From (\ref{dw_exchange2}) we derive
\begin{equation}
e^{i\Theta_i} \ {\rm id}_V = e^{i 2\pi N_i s_\alpha} \ {\rm id}_V = \rho(e_i)^{N_i}
\end{equation}
We then rewrite the right hand side as
\begin{eqnarray}
\rho(e_i)^{N_i} &=& \prod_{n=0}^{N_i-1} (\rho((n+1)e_i)^{-1} \rho(e_i) \rho(n e_i)) \nonumber \\
&=& \prod_{n=0}^{N_i-1}\chi_{e_i}(e_i, ne_i) \ {\rm id}_V
\end{eqnarray}
where the second line follows from equation (\ref{projective}). We conclude that the invariant $\Theta_i$ is given by
\begin{equation}
\exp(i\Theta_i )  = \prod_{n=0}^{N_i-1}\chi_{e_i}(e_i, ne_i) \label{theta_expression1}
\end{equation}
Similarly, we can obtain expressions for $\Theta_{ij}, \Theta_{ijk}$ using (\ref{dw_braiding}),
\begin{align*}
e^{i\Theta_{ij}} \ {\rm id}_V\otimes {\rm id}_W & = \rho(e_j)^{N^{ij}}\otimes \mu(e_i)^{N^{ij}}, \nonumber \\
 e^{i\Theta_{ijk}} {\rm id}_V & = \rho(e_k)^{-1}\rho(e_j)^{-1}\rho(e_k)\rho(e_j),
\end{align*}
which can then be related to $\chi$ using (\ref{projective}):
\begin{align}
\exp( i\Theta_{ij}) & = \prod_{n=1}^{N^{ij}}\chi_{e_i}(e_j, ne_j)\chi_{e_j}(e_i, ne_i) \nonumber \\
\exp(i \Theta_{ijk} )& =  \frac{\chi_{e_i}(e_k, e_j)}{\chi_{e_i}(e_j, e_k)} \label{theta_expression2}
\end{align}
Equations (\ref{theta_expression1}, \ref{theta_expression2}) are the formulas we seek, where
$\chi$ is defined by (\ref{slant1}).

The properties of $\Theta_i, \Theta_{ij}, \Theta_{ijk}$ that were derived in Sec.~\ref{sec:2dnonabelian} from more general considerations are manifest in the above expressions. We can see that $\Theta_i, \Theta_{ij}, \Theta_{ijk}$  only depend on $i,j,k$ and not on the choice of vortices $\alpha, \beta, \gamma$,
since the representations $\rho,\mu,\nu$ do not appear in the final expressions (\ref{theta_expression1},\ref{theta_expression2}).
In addition, it is easy to verify that these formulas are invariant under the change $\omega \rightarrow \omega \cdot \nu$ if $\nu$
is a coboundary (see Appendix \ref{sec:appd_cohomology} for the definition of a coboundary). This is to be expected, since
two cocycles that differ by a coboundary are known to define the same Dijkgraaf-Witten model, and
therefore must give the same values for the invariants.

\subsection{3D topological invariants}

\begin{figure}[b]
\includegraphics{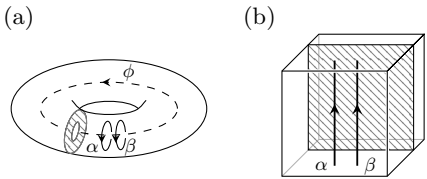}
\caption{(a) A thickened torus $\mathbb T^2\times[0,1]$ with a flux $\phi$ threading the inner hole. (b) The thickened torus drawn as a cube with the top and bottom faces as well as the front and back faces identified. } \label{fig_dr}
\end{figure}
\subsubsection{Dimensional reduction}
Our approach for computing the 3D invariants is based on dimensional reduction: we derive a relationship between
vortex loop statistics in 3D Dijkgraaf-Witten models and vortex statistics
in 2D Dijkgraaf-Witten models, and then we analyze the latter using previously known 2D results.
A similar dimensional reduction approach was used in Ref.~\onlinecite{threeloop}. The derivation we discuss here is
more general than that of Ref.~\onlinecite{threeloop} in some ways and less general in other ways. It is more
general because it applies even if the vortex loop statistics are non-Abelian, but it is also less general
because it is only valid for Dijkgraaf-Witten models, while
Ref.~\onlinecite{threeloop} derived a dimensional reduction formula without restricting to these exactly soluble systems.

To begin, consider a Dijkgraaf-Witten model associated with a group $G=\prod_{i=1}^K\mathbb Z_{N_i}$ and a
4-cocycle $\omega$. Let us define this model on a thickened 2D torus, i.e., the manifold ${\mathbb T}^2\times[0,1]$
(Fig.~\ref{fig_dr}). Consider a state consisting of two loop excitations $\alpha, \beta$ which wind around the inner hole
of the torus, and suppose that there is a flux
\begin{equation}
\phi = (\frac{2\pi}{N_1} a_1,...,\frac{2\pi}{N_k} a_k) \label{adef}
\end{equation}
threading the inner hole. This geometry is equivalent to a standard three-loop setup in which $\alpha, \beta$ are
linked with another loop carrying flux $\phi$. Our task is to find the unitary transformation associated
with braiding $\alpha$ around $\beta$ in the presence of the flux $\phi$.

To this end, we redraw the thickened torus ${\mathbb T}^2\times[0,1]$ as a cube whose top and bottom faces are
identified as well as its front and back faces. In this representation, the shaded square in Fig.~\ref{fig_dr}(b) corresponds to the shaded annulus
in Fig.~\ref{fig_dr}(a) and the left and right faces of the cube correspond to the inner and outer surfaces of the thickened
torus. The flux loops $\alpha,\beta$ can be drawn as lines connecting the top and bottom faces of the cube.

To proceed further we make use of a special property of Dijkgraaf-Witten models: these models can be defined for any
triangulation of space-time, and their properties do not depend on the choice of triangulation. Therefore, we
are free to triangulate our cube however we like and it won't affect the braiding statistics between
$\alpha$ and $\beta$. Making use of this freedom, we consider a triangulation with translational symmetry in the $x$, $y$ and $z$
directions. More specifically, we consider triangulation in which there is only one unit cell in the $z$ direction, but
many unit cells in the $x$ and $y$ direction. With this choice of triangulation, the cube can be viewed as a 2D system.
Furthermore, the braiding process involving the vortex loops $\alpha$ and $\beta$ can be viewed as a process involving
two vortices in this effective 2D system.

By the same argument as in the supplementary material of Ref.~\onlinecite{threeloop}, one can show that this
effective 2D system is identical to a 2D Dijkgraaf-Witten model with group $G$ and a 3-cocycle
\begin{equation}
\chi_a(b,c,d) = \frac{\omega(b,a,c,d)\omega(b,c,d,a)}{\omega(a,b,c,d)\omega(b,c,a,d)} \label{slant2}
\end{equation}
where $a =(a_1,...,a_K)$ is defined in terms of $\phi$ as in equation (\ref{adef}). Here, $\chi_a$ is known as the
{\it slant product} of $\omega$. One can check that $\chi_a$ is indeed a 3-cocycle for any choice of $a$.

Putting this all together, we conclude that the vortex loop statistics for a 3D Dijkgraaf-Witten model with 4-cocycle
$\omega$ and with a base loop with flux $\phi$, are identical to the vortex statistics in a 2D Dijkgraaf-Witten model with
3-cocycle $\chi_a$.

\subsubsection{Explicit formulas for the invariants}
We are now ready to compute the invariants $\Theta_{i,l},\Theta_{ij,l}, \Theta_{ijk, l}$ for a 3D Dijkgraaf-Witten model with
group $G = \prod_{i=1}^K \mathbb Z_{N_i}$ and 4-cocycle $\omega$.

Let $\alpha, \beta, \gamma$ be three vortex loops carrying unit flux $2\pi e_i/N_i$, $2\pi e_j/N_j$ and $2\pi e_k/N_k$
respectively, and suppose that all three are linked with another loop $\sigma$ carrying unit flux $2\pi e_l/ N_l$.
To compute $\Theta_{i,l},\Theta_{ij,l}, \Theta_{ijk, l}$, we need to analyze various braiding processes involving
$\alpha, \beta, \gamma$. We can accomplish this task with the help of the
dimensional reduction results of the previous section: according to those results, the braiding statistics of the vortex
loops $\alpha, \beta, \gamma$ in a 3D Dijkgraaf-Witten model with cocycle $\omega$ are
identical to the braiding statistics of vortices in a 2D Dijkgraaf-Witten model with 3-cocycle $\chi_{e_l}$ (\ref{slant2}).
Therefore the 3D invariants $\Theta_{i,l}, \Theta_{ij,l}, \Theta_{ijk, l}$ can be obtained from the corresponding
2D invariants (\ref{theta_expression1}, \ref{theta_expression2}) simply by substituting $\chi_{e_l}(a,b,c)$ for $\omega(a,b,c)$.
In this way, we obtain
\begin{align}
\exp(i\Theta_{i,l} ) & = \prod_{n=1}^{N_i}\chi_{e_l,e_i}(e_i, ne_i) \nonumber\\
\exp( i\Theta_{ij,l}) & = \prod_{n=1}^{N^{ij}}\chi_{e_l, e_i}(e_j, ne_j)\chi_{e_l, e_j}(e_i, ne_i)\nonumber  \\
\exp(i \Theta_{ijk,l} )& =  \frac{\chi_{e_l, e_i}(e_k, e_j)}{\chi_{e_l, e_i}(e_j, e_k)} \label{theta_expression_3D}
\end{align}
where $\chi_{e_l,e_i}$ is defined as
\begin{equation}
\chi_{e_l,e_i}(b,c) = \frac{\chi_{e_l}(e_i,b,c) \chi_{e_l}(b,c,e_i)}{\chi_{e_l}(b,e_i,c)}
\end{equation}
and where $\chi_{e_l}$ is defined in (\ref{slant2}).

\section{Showing the invariants distinguish all Dijkgraaf-Witten models}
\label{sec:distinguishingDW}

In this section, we show that the invariants take different values for each of the Dijkgraaf-Witten models with group $G$. This result has two
implications: (i) each of the Dijkgraaf-Witten models belongs to a distinct phase and (ii) the invariants
can distinguish these phases.

\subsection{2D case}
We now show that the invariants $\Theta_i, \Theta_{ij}, \Theta_{ijk}$ can distinguish all the 2D Dijkgraaf-Witten models corresponding to
the group $G = \prod_{i=1}^K Z_{N_i}$ (A similar result was obtained previously in Ref.~\onlinecite{zaletel}). Our proof is based on a counting argument: let $\mathcal N^{\rm DW}_{2D}$ be the number of
2D Dijkgraaf-Witten models with group $G$, let $\mathcal N^{\rm phase}_{2D}$ be the number of phases of 2D Dijkgraaf-Witten models with group $G$, and let $\mathcal N^{\Theta}_{2D}$ be the number of distinct values that the invariants take over all 2D Dijkgraaf-Witten models with group $G$. Clearly we must have
\begin{equation}
\mathcal N^{\Theta}_{2D} \leq \mathcal N^{\rm phase}_{2D} \leq \mathcal N^{\rm DW}_{2D}.
\end{equation}
What we will prove is the opposite inequality:
\begin{equation}
\mathcal N^{\Theta}_{2D} \geq \mathcal N^{\rm DW}_{2D}
\label{2dbound}
\end{equation}
It will then follow that $\mathcal N^{\Theta}_{2D} = \mathcal N^{\rm phase}_{2D} = \mathcal N^{\rm DW}_{2D}$, which shows that each of the Dijkgraaf-Witten models belongs to a distinct phase and the invariants can distinguish all the phases (according to the definition of phases given in Sec.~\ref{sec:model_phase}).

To prove (\ref{2dbound}) we consider the set of $3$-cocycles of the form
\begin{align}
\omega(a,b,c) = & \exp\{i 2\pi \sum_{ij} \frac{P_{ij}}{N_iN_j} a_i(b_j + c_j - [b_j+c_j])\} \nonumber\\
 & \times \exp\{i2 \pi \sum_{ijk} \frac{Q_{ijk}}{N_{ijk}} a_ib_jc_k\},
\end{align}
where $P_{ij}$ is an integer matrix  and $Q_{ijk}$ is an integer tensor with $Q_{ijk} = 0$ if $i,j,k$ are not all distinct. (The latter restriction on $Q_{ijk}$ is not essential, and we include it only to simplify some of the formulas that follow). Here, the symbol $[b_j+c_j]$ is defined as the residue of $b_j+c_j$ modulo $N_j$ with values taken in the range $0,...,N_j-1$. (One can verify that $\omega$ obeys the $3$-cocycle condition (\ref{3cocyclecond}) with straightforward algebra). Inserting the cocycle $\omega$ into the expressions in (\ref{theta_expression1},\ref{theta_expression2}), one immediately obtains
\begin{subequations}
\begin{align}
\Theta_i &= \frac{2\pi}{ N_i}P_{ii} \label{theta_value_2d1}\\
\Theta_{ij} &= \frac{2\pi}{N_{ij}} (P_{ij}+P_{ji}) \label{theta_value_2d2}\\
\Theta_{ijk} & = -\frac{2\pi}{N_{ijk}}(Q_{ijk}+Q_{jki}+Q_{kij}-Q_{jik}-Q_{ikj}-Q_{kji}) \label{theta_value_2d3}
\end{align}
\end{subequations}
From the above formulas, we see that the invariants $\Theta_{ij}$ can take on $N_{ij}$ different values as $P_{ij}$ ranges over all integer matrices. Similarly, $\Theta_i$ can take on $N_i$ different values while $\Theta_{ijk}$  can take on
$N_{ijk}$ different values. Furthermore, the values of $\Theta_{i}$, and $\Theta_{ij}$ with $i < j$, and $\Theta_{ijk}$ with $i < j < k$, can be varied independently from one another. Therefore, we have the lower bound
\begin{equation}
\mathcal N^{\Theta}_{2D} \geq \prod_{i} N_i \prod_{i<j} N_{ij} \prod_{i<j<k} N_{ijk}.
\label{2dthetabound}
\end{equation}
At the same time, we know that the Dijkgraaf-Witten models are parameterized by elements of the group
\begin{equation}
H^3[G, U(1)] =  \prod_{ i} \mathbb Z_{N_i}\prod_{i < j} \mathbb Z_{N_{ij}}\prod_{i < j <k} \mathbb Z_{N_{ijk}} .
\end{equation}
so that
\begin{equation}
\mathcal N^{\rm DW}_{2D} = |H^3(G, U(1))| = \prod_{i} N_i \prod_{i<j} N_{ij} \prod_{i<j<k} N_{ijk}
\label{2ddw}
\end{equation}
Combining (\ref{2dthetabound}) and (\ref{2ddw}) gives the desired inequality (\ref{2dbound}) and proves the result.

\subsection{3D case}

We now show that the invariants $\Theta_{i,l},\Theta_{ij,l}, \Theta_{ijk, l}$ can distinguish all the 3D Dijkgraaf-Witten models corresponding to the group $G = \prod_{i=1}^K Z_{N_i}$. The argument closely follows the 2D case: let $\mathcal N^{\rm DW}_{3D}$ be the number of 3D Dijkgraaf-Witten models with group $G$, let $\mathcal N^{\rm phase}_{3D}$ be the number of phases of 3D Dijkgraaf-Witten models with group $G$, and let $\mathcal N^{\Theta}_{3D}$ be the number of distinct values that the invariants take over all 3D Dijkgraaf-Witten models with group $G$. Clearly, we have
\begin{equation}
\mathcal N^{\Theta}_{3D} \leq \mathcal N^{\rm phase}_{3D} \leq \mathcal N^{\rm DW}_{3D}
\end{equation}
What we will show is that
\begin{equation}
\mathcal N^{\Theta}_{3D} \geq \mathcal N^{\rm DW}_{3D}.
\label{3dbound}
\end{equation}
It will then follow that $\mathcal N^{\Theta}_{3D} = \mathcal N^{\rm phase}_{3D} = \mathcal N^{\rm DW}_{3D}$, which shows that each of the Dijkgraaf-Witten models belongs a distinct phase and the invariants distinguish all the phases (according to the definition of phases given in Sec.~\ref{sec:model_phase}).

To prove (\ref{3dbound}), we consider the set of 4-cocycles of the form
\begin{align}
\omega(a,b,c,d) = & \exp\{i 2\pi\sum_{ijk} \frac{ M_{ijk}}{N_{ij}N_k} a_ib_j(c_k+d_k-[c_k+d_k])\} \nonumber\\
 & \times \exp\{i2\pi \sum_{ijkl} \frac{L_{ijkl}}{N_{ijkl}} a_ib_jc_kd_l\}.
\label{omega3d}
\end{align}
where $M_{ijk}$ is an arbitrary integer tensor and $L_{ijkl}$ is an integer tensor with $L_{ijkl} = 0$ if $i,j,k$ are not all distinct. (As in the 2D case, it is simple to verify that $\omega$ obeys the 4-cocycle condition (\ref{4cocyclecond})).
Inserting the 4-cocycle $\omega$ into the expressions in (\ref{theta_expression_3D}), we obtain
\begin{subequations}
\begin{align}
\Theta_{i,l} &= \frac{2\pi}{N_{il}}(M_{ili}-M_{lii}) \label{theta_value_3d1}\\
\Theta_{ij,l} &= \frac{2\pi N^{ij}}{N_{il} N_j}(M_{ilj}- M_{lij}) + \frac{2\pi N^{ij}}{N_{jl} N_i}(M_{jli}- M_{lji})\label{theta_value_3d2} \\
\Theta_{ijk,l} & =-\frac{2\pi}{N_{ijkl}} \sum_{\hat p} {\rm sgn}(\hat p) L_{\hat p(i)\hat p(j)\hat p(k)\hat p(l)} \label{theta_value_3d3}
\end{align}
\end{subequations}
where $\hat p$ is a permutation of $i,j,k,l$ and ${\rm sgn}(\hat p)=\pm 1$ is the parity of $\hat p$. From the above formulas, we see that different choices of $M$ and $L$ give different values of $\Theta_{i,l}, \Theta_{ij,l}$ and $\Theta_{ijk,l}$. More precisely, it can be shown that as $M$ and $L$ range over the set of allowed integer tensors, the invariants $\Theta_{i,l}, \Theta_{ij,l}$ and $\Theta_{ijk,l}$ take on at least $\prod_{i<l} (N_{il})^2 \prod_{i<j<l} (N_{ijl})^2 \prod_{i<j<k<l} N_{ijkl}$ different values (this counting is done in Appendix \ref{sec:appd_counting}). Therefore we have the lower bound
\begin{equation}
\mathcal N^{\Theta}_{3D} \geq \prod_{i<l} (N_{il})^2 \prod_{i<j<l} (N_{ijl})^2 \prod_{i<j<k<l} N_{ijkl}.
\label{3dthetabound}
\end{equation}

On the other hand, we know that the Dijkgraaf-Witten models are parameterized by elements of the group
\begin{equation}
H^4[G, U(1)] =  \prod_{i < j} \left(\mathbb Z_{N_{ij}}\right)^2\prod_{ i < j <k} \left(\mathbb Z_{N_{ijk}}\right)^2 \prod_{ i < j <k<l} \mathbb Z_{N_{ijkl}}.
\end{equation}
so that
\begin{equation}
\mathcal N^{\rm DW}_{2D} = \prod_{i<l} (N_{il})^2 \prod_{i<j<l} (N_{ijl})^2 \prod_{i<j<k<l} N_{ijkl}.
\label{3ddw}
\end{equation}
Combining (\ref{3dthetabound}) and (\ref{3ddw}) gives the desired inequality (\ref{3dbound}) and proves the result.

\section{General constraints on the invariants}

\label{sec:generalconstraint}
In this section, we derive general constraints on the invariants that hold for any gauge theory with group $G = \prod_{i=1}^K \mathbb Z_{N_i}$. In the next section, we will discuss whether these constraints are complete, i.e. whether any solution to these constraints
can be realized by an appropriate gauge theory.

\subsection{2D Abelian case}
\label{sec:2dabelian_constraint}
Let us start with the case of 2D gauge theories with gauge group $G = \prod_{i=1}^K \mathbb Z_{N_i}$ and Abelian braiding statistics. According to Sec.~\ref{sec:2dabelian}, there are two invariants $\Theta_i$ and $\Theta_{ij}$ in this case. We will now argue that $\Theta_i$ and $\Theta_{ij}$ must satisfy the following general constraints:
\begin{subequations}
\begin{align}
\Theta_{ii} & = 2\Theta_i, \label{2d_exch}\\
\Theta_{ij} & = \Theta_{ji}, \label{2d_sym}\\
N_{ij}\Theta_{ij} & = 0,\label{2d_quant}\\
N_i\Theta_i & =0, \label{2d_quant_ex}
\end{align}
\end{subequations}
where all equations are defined modulo $2\pi$. To prove this statement, we first recall the following general properties of Abelian braiding statistics:
\begin{subequations}
\begin{align}
\theta_{\alpha\beta} & = \theta_{\beta\alpha}, \label{symmetry_2d}\\
\theta_{\alpha\alpha} &= 2\theta_{\alpha}\label{exchange_2d} \\
\theta_{\alpha(\beta_1 \times \beta_2)} & = \theta_{\alpha\beta_1} + \theta_{\alpha\beta_2}, \label{linearity_2d} \\
\theta_{(\alpha \times \beta)} & = \theta_\alpha + \theta_\beta + \theta_{\alpha\beta}.\label{linearity_2d_exch}
\end{align}
\end{subequations}
Here $\theta_{\alpha \beta}$ denotes the mutual statistics of $\alpha, \beta$, while $\theta_\alpha$ denotes the exchange statistics of $\alpha$ and $\alpha \times \beta$ denotes the excitation created by fusing together $\alpha$ and $\beta$. Each of these identities follow from simple physical arguments. The symmetry relation (\ref{symmetry_2d}) comes from the fact that braiding $\alpha$ around $\beta$ is topologically equivalent to braiding $\beta$ around $\alpha$. The relation (\ref{exchange_2d}) follows immediately from the definition of exchange statistics. The linearity relation (\ref{linearity_2d}) comes from the fact that fusing $\beta_1$ and $\beta_2$ must commute with braiding $\alpha$ around them. The other linearity relation (\ref{linearity_2d_exch}) has a similar flavor.

We can see that equations (\ref{2d_exch}) and (\ref{2d_sym}) follow immediately from these general constraints. To prove (\ref{2d_quant}), consider braiding a vortex $\alpha$ that carries a unit flux $\frac{2\pi}{N_i}e_i$ around $N_j$ identical vortices $\beta$, with each $\beta$ carrying a unit flux $\frac{2\pi}{N_j}e_j$. Clearly the associated statistical phase is $N_j\theta_{\alpha\beta}$. At the same time, according to the linearity relation (\ref{linearity_2d}), this statistical phase is equal to the phase associated with braiding $\alpha$ around the fusion product
of all the $\beta$ vortices.  Since the total flux of $N_j$ $\beta$ vortices is zero, they fuse to a charge. It then follows from the Aharanov-Bohm law that the latter quantity
is a multiple of $\frac{2\pi}{N_i}$. We conclude that
\begin{equation}
N_j\theta_{\alpha\beta} = \frac{2\pi}{N_i}\times \text{integer}. \nonumber
\end{equation}
Equation (\ref{2d_quant}) follows immediately using $N_{ij}N^{ij}=N_iN_j$ together with the definition (\ref{2d_Theta}) of $\Theta_{ij}$.

To prove (\ref{2d_quant_ex}), imagine exchanging a set of $N_i$ $\alpha$'s with another set of $N_i$ $\alpha$'s. According to the linearity relation (\ref{linearity_2d_exch}) and the relation (\ref{exchange_2d}), the associated exchange statistical phase is $N_i^2\theta_\alpha=N_i\Theta_i$. However, this phase must be a multiple of $2\pi$ since $N_i$ $\alpha$'s fuse to a charge which is a boson. Thus, equation (\ref{2d_quant_ex}) must hold.

\subsection{2D general case}
\label{sec:2dnonabelian_constraint}
We now consider the case of general 2D gauge theories with gauge group $G = \prod_{i=1}^K \mathbb Z_{N_i}$. There are three invariants,
$\Theta_i, \Theta_{ij}, \Theta_{ijk}$ in this case. We will now argue that $\Theta_i, \Theta_{ij}$ again satisfy the constraints (\ref{2d_exch})-(\ref{2d_quant_ex}), as in the Abelian case. In addition, the third invariant $\Theta_{ijk}$ satisfies
\begin{subequations}
\begin{align}
\Theta_{ijk}  & = {\rm sgn}(\hat p) \Theta_{\hat p(i)\hat p(j)\hat p(k)}, \label{2d_antisym_ijk}\\
\Theta_{iij} & = 0, \label{2d_antisym_ijk2}\\
N_{ijk} \Theta_{ijk} & = 0, \label{2d_quant_ijk}
\end{align}
\end{subequations}
where $\hat p$ is a permutation of the indices $i,j,k$ and ${\rm sgn}(\hat p)=\pm 1$ is its parity and again all the equations are defined modulo $2\pi$. We note that the constraint (\ref{2d_antisym_ijk}) tells us that $\Theta_{ijk}$ is fully antisymmetric modulo $2\pi$,
while (\ref{2d_antisym_ijk2}) is a stronger constraint on $\Theta_{iij}$ than its antisymmetry, since the antisymmetry only requires that
$2\Theta_{iij}=0 \ ({\rm mod} \ 2\pi)$.

\begin{figure}
\includegraphics{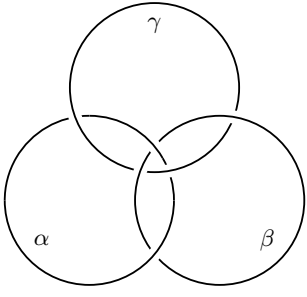}
\caption{A Borromean ring obtained by closing up the trajectories in Fig.~\ref{fig_spacetimetraj}(b). }
\label{fig_borromean}
\end{figure}

We first prove the above constraints for $\Theta_{ijk}$; afterwards we will prove (\ref{2d_exch})-(\ref{2d_quant_ex}). To prove that $\Theta_{ijk}$ is fully antisymmetric, i.e. (\ref{2d_antisym_ijk}), we consider the space-time trajectories (Fig.~\ref{fig_spacetimetraj}b) of the vortices in the braiding process associated with $\Theta_{ijk}$. Since the unitary transformation associated with the braiding process is Abelian, we can close up the space-time trajectories in Fig.\ref{fig_spacetimetraj}b, so that they form the Borromean ring in Fig.\ref{fig_borromean}. The closed-up trajectories are associated with the following process: we first create three particle-antiparticle pairs $\alpha,\bar\alpha$, $\beta,\bar\beta$ and $\gamma,\bar\gamma$ out of the vacuum,  where $\alpha,\beta,\gamma$ carry unit flux $\frac{2\pi}{N_i}e_i,\frac{2\pi}{N_j}e_j,\frac{2\pi}{N_k}e_k$ respectively, then braid $\alpha,\beta,\gamma$ in the way that leads to the phase $\Theta_{ijk}$, and finally annihilate the three pairs to return to the vacuum. The fact that we can annihilate the particles at the end of the process is guaranteed by the fact that the braiding results in a pure phase, otherwise the particle-antiparticle need not be in the vacuum fusion channel after the braiding.

With this picture in mind, we can see that $\Theta_{ijk}$ is equal to the phase associated with the Borromean ring space-time trajectories. Since the Borromean ring is cyclically symmetric, we deduce that $\Theta_{ijk}=\Theta_{jki}=\Theta_{kij}$. It is also not hard to see that reversing the braiding process associated with $\Theta_{ijk}$ gives rise to the phase $\Theta_{ikj}$. So, $\Theta_{ijk}=-\Theta_{ikj}$. Putting these relations together, we see $\Theta_{ijk}$ is fully antisymmetric.

To prove (\ref{2d_antisym_ijk2}), we make use of the result in Appendix \ref{sec:appd_Raa} which shows that braiding a vortex $\alpha$ around another $\alpha$ gives only a pure phase. Consider three vortices $\alpha,\alpha, \beta$ with $\phi_\alpha=\frac{2\pi}{N_i}e_i$ and $\phi_\beta = \frac{2\pi}{N_j}e_j$ and imagine the braiding process associated with $\Theta_{iij}$: $\alpha$ is braided around the other $\alpha$, then around $\beta$, then around the other $\alpha$ in the opposite direction, and finally around $\beta$ in the opposite directions. In this four-step process, we can switch the order of the second and third step since the third step gives only a pure phase. Thus, it is obvious that this four-step process neither changes the state of the system, nor leads to any Abelian phase. Hence, equation (\ref{2d_antisym_ijk2}) holds.

\begin{figure}
\includegraphics{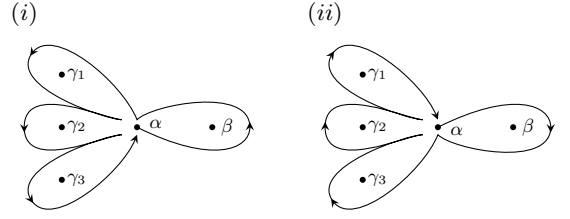}
\caption{ Thought experiment to prove the constraint (\ref{2d_quant_ijk}) ($N_k=3$ is taken for illustration).  For clarity, we split the composite braiding into two steps ($i$) and ($ii$). In the thought experiment, ($ii$) follows immediately after ($i$). } \label{fig_thoughtexp1}
\end{figure}

To prove (\ref{2d_quant_ijk}), we consider a collection of excitations $\alpha, \beta, \gamma_1, \dots, \gamma_{N_k}$, where $\phi_\alpha= \frac{2\pi}{N_i}e_i$, $\phi_\beta= \frac{2\pi}{N_j}e_j$ and $\phi_{\gamma_t}= \frac{2\pi}{N_k}e_k$ for any $t=1, \dots, N_k$.  We consider a composite braiding process shown in Fig.~\ref{fig_thoughtexp1}: first we braid $\alpha$ around $\beta$ in the counterclockwise direction, then sequentially round $\gamma_1, \dots, \gamma_{N_k}$ in the counterclockwise direction, then around $\beta$ in the clockwise direction, and sequentially around $\gamma_{N_k}, \dots, \gamma_1$ in the clockwise direction. This braiding process can be described by a product of operators
\begin{equation}
B \equiv B_{\alpha\gamma_1}^{-1}\dots B_{\alpha\gamma_{N_k}}^{-1}B_{\alpha\beta}^{-1} B_{\alpha\gamma_{N_k}}\dots B_{\alpha\gamma_1}B_{\alpha\beta} \nonumber,
\end{equation}
where $B_{\alpha\beta}, B_{\alpha\gamma_1}, \dots, B_{\alpha\gamma_{N_k}}$ are the operators associated with braiding $\alpha$ around $\beta, \gamma_1, \dots, \gamma_{N_k}$ respectively. Now according to the definition of $\Theta_{ijk}$, we have $B_{\alpha\gamma_t}^{-1}B_{\alpha\beta}^{-1}B_{\alpha\gamma_{t}}B_{\alpha\beta}=e^{i\Theta_{ijk}} {\hat I}$ for any $t=1, \dots, N_k$. It then follows that
\begin{equation}
B = e^{i N_k \Theta_{ijk}} {\hat I}
\end{equation}
where $\hat I$ is the identity operator. On the other hand, braiding $\alpha$ around $\gamma_1, \dots, \gamma_{N_k}$ in sequence is equivalent to braiding $\alpha$ around the fusion product of $\gamma_1, \dots, \gamma_{N_k}$. Since $\gamma_1, \dots, \gamma_{N_k}$ fuse to some charge $q$, we derive
\begin{equation}
B = B_{\alpha q}^{-1}B_{\alpha\beta}^{-1}B_{\alpha q}B_{\alpha\beta},
\end{equation}
where $B_{\alpha q}$ denotes the operator associated with the braiding of $\alpha$ around $q$. Now, by the Aharonov-Bohm formula, we know
that $B_{\alpha q}$ is a pure phase, which implies that $B$ is just the identity operator, $B = \hat I$. Therefore, $e^{i N_k\Theta_{ijk}}=1$, i.e., $N_{k}\Theta_{ijk}=0$. Similarly, we can show $N_i \Theta_{ijk} = N_j\Theta_{ijk}=0$. Putting this together, we derive the constraint (\ref{2d_quant_ijk}).

Next, we prove the constraints (\ref{2d_exch})-(\ref{2d_quant_ex}) in the non-Abelian case. The constraint (\ref{2d_exch}) follows immediately from Appendix \ref{sec:appd_Raa} while (\ref{2d_sym}) is obvious. To prove (\ref{2d_quant}), we consider a vortex $\alpha$ carrying a flux $ \frac{2\pi}{N_i}e_i$, together with $N_j$ vortices $\beta_1, \dots, \beta_{N_j}$ all carrying a flux $ \frac{2\pi}{N_j}e_j$. Imagine $\alpha$ is braided around $\beta_1$ for $N^{ij}$ times, then around $\beta_2$ for $N^{ij}$ times, and so on. The result is a total phase $N_j\Theta_{ij}$. This sequence of braiding processes can be described by a product of operators
\begin{equation}
B' =B_{\alpha\beta_{N_j}}^{N^{ij}}\cdots B_{\alpha\beta_2}^{N^{ij}}B_{\alpha\beta_1}^{N^{ij}} \label{number}
\end{equation}
where $B_{\alpha\beta_t}$ represents the operator associated with braiding $\alpha$ around $\beta_{t}$ once. Any two operators $B_{\alpha\beta_t}$ and $B_{\alpha\beta_s}$ commute, because the commutator $B_{\alpha\beta_t}^{-1}B_{\alpha\beta_s}^{-1}B_{\alpha\beta_t}B_{\alpha\beta_s} = e^{i \Theta_{ijj}} $ and $\Theta_{ijj}=0$ according to (\ref{2d_antisym_ijk2}). Therefore, the operator $B'$ can be rewritten as
\begin{equation}
B' = (B_{\alpha\beta_{N_j}}\cdots B_{\alpha\beta_2}B_{\alpha\beta_1} )^{N^{ij}},
\end{equation}
which means $B'$ is equivalent to braiding $\alpha$ around $\beta_1, \dots, \beta_{N_j}$ as a whole for $N^{ij}$ times. However, the vortices $\beta_1, \dots, \beta_{N_j}$ fuse to a pure charge, and when $\alpha$ is braided around any charge for $N^{ij}$ times, the result is no phase at all. Therefore, we obtain $N_j\Theta_{ij}=0$. Similarly, one can show that $N_i\Theta_{ij}=0$. Putting this together, we
derive the constraint (\ref{2d_quant}).

Finally, to prove the constraint (\ref{2d_quant_ex}), we use the diagrammatical representation of the topological spin\cite{kitaev06}
\begin{equation}
\includegraphics{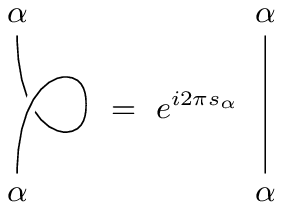}
\end{equation}
Let us imagine $N_i$ identical $\alpha$'s, which should fuse to some charge $q$. Consider the following diagram
\begin{equation}
\includegraphics{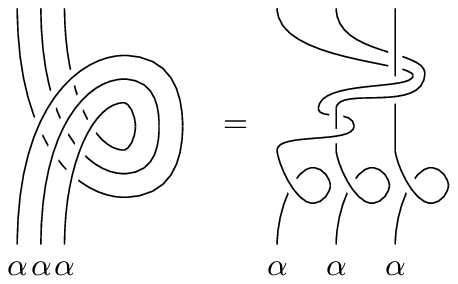}
\end{equation}
where the case $N_i=3$ is shown for simplicity. The left hand side equals the topological spin $e^{i2\pi s_q}=1$, while the right hand side equals
\begin{equation}
e^{i 2\pi N_i  s_\alpha + \sum_{n=0}^{N_i-1} i4\pi n s_\alpha} = e^{i2\pi N_i^2 s_\alpha} = e^{i N_i \Theta_{i}},
\end{equation}
where the result of Appendix \ref{sec:appd_Raa} is used. We conclude that $N_i\Theta_{i}=0$ modulo $2\pi$, which proves the constraint
(\ref{2d_quant_ex}).

%Consider the following graphics in Fig.~\ref{fig_topospin}. One the leftmost side, the graph represents the topological spin $\theta_q$ to the power of $N_i$, which is 0. The rightmost graph gives
%\begin{equation}
%(\theta_\alpha)^{N_i^2} \exp\left(i2\sum_{n=1}^{N_i-1} n\Theta_i\right) = e^{i N_i^2\Theta_i}. \label{appdf_topo}
%\end{equation}
%Therefore, we have $N_i^2\Theta_i = 0$.  According to the constraint (\ref{2d_quant}), we also have $2N_i \Theta_i =0$. Combining the two, we have
%\begin{equation}
%{\rm gcd}(2, N_i)\times N_i \Theta_i =0.
%\end{equation}
%Therefore, if $N_i$ is odd, we have $N_i\Theta_i=0$.

%\begin{figure*}
%\includegraphics[scale=1.3]{fig_topospin.eps}
%\caption{Graphical proof of $N_i^2\Theta_i=0$. Leftmost represents the topological spin $\theta_q$ to the $N_i$th power ($N_i=3$). Rightmost represents the quantity in (\ref{appdf_topo}). }\label{fig_topospin}
%\end{figure*}

\subsection{3D Abelian case}
\label{sec:3dabelian_constraint}
We now consider the case of 3D gauge theories with gauge group $G = \prod_{i=1}^K \mathbb Z_{N_i}$ and Abelian braiding statistics.
In this case, there are two invariants $\Theta_{i,l}$ and $\Theta_{ij,l}$. We will argue that these invariants must satisfy the following general constraints:
\begin{subequations}
\label{3dconstraints}
\begin{align}
&\Theta_{ii, l}   = 2 \Theta_{i,l}, \label{3dconstraint_exch}\\
&\Theta_{ij,l}  =\Theta_{ji, l},\label{3dconstraint_sym}\\
&N_{ijl}\Theta_{ij,l}  = 0, \label{3dconstraint_quant}\\
&N_{il}\Theta_{i,l}   = 0,\label{3dconstraint_quant_ex}\\
&\frac{N^{ijl}}{N^{ij}} \Theta_{ij,l} +\frac{N^{ijl}}{N^{jl}} \Theta_{jl,i} +\frac{N^{ijl}}{N^{li}} \Theta_{li,j}=0, \label{3dconstraint_cyclic}\\
&\Theta_{i,l}\frac{N^{il}}{N_i} + \Theta_{il,i}  =0, \label{3dconstraint_cyclic2}\\
&\Theta_{i,i}   =0. \ \ \ \ \text{(conjectured)}\label{3dconstraint_cyclic3}
\end{align}
\end{subequations}
We note that the above constraints are a generalization of those derived in Ref.~\onlinecite{threeloop}.

\begin{figure}
\includegraphics{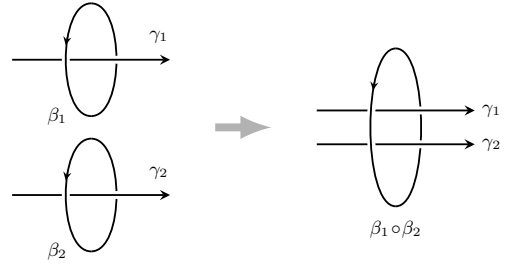}
\caption{Fusion of two loops $\beta_1$  and $\beta_2$ that carry the same amount of flux $\phi_{\beta_1}=\phi_{\beta_2}$ and that are linked to different base loops. We denote this type of fusion by $\beta_1\circ\beta_2$. This is different notation from Ref.~\onlinecite{threeloop}, where this type of fusion was denoted by $\beta_1 \oplus \beta_2$. } \label{fig_fusionb}
\end{figure}

To prove these constraints, we make use of the following general properties of Abelian three-loop statistics:
\begin{subequations}
\begin{align}
\theta_{\alpha\alpha,c} & = 2\theta_{\alpha,c}, \label{exch3d}\\
\theta_{\alpha\beta,c} & = \theta_{\beta\alpha,c}, \label{symmetry3d} \\
\theta_{\alpha(\beta_1 \times \beta_2),c} & = \theta_{\alpha\beta_1, c} + \theta_{\alpha\beta_2, c}, \label{linearity3d1} \\
\theta_{(\alpha \times \beta),c} & = \theta_{\alpha,c} + \theta_{\beta, c} + \theta_{\alpha\beta,c},  \label{linearity3d_ex1}\\
\theta_{(\alpha_1\circ\alpha_2)(\beta_1\circ\beta_2), (c_1+c_2)} & = \theta_{\alpha_1\beta_1,c_1} + \theta_{\alpha_2\beta_2,c_2}, \label{linearity3d2}\\
\theta_{\alpha\circ\beta,c_1+c_2} & = \theta_{\alpha,c_1} + \theta_{\beta,c_2}, \label{linearity3d_ex2}
\end{align}
\end{subequations}
Note that these equations involve two types of fusions of loops, i.e. the ``$\times$'' and ``$\circ$'' fusions, shown in Fig.~\ref{fig_fusion} and Fig.~\ref{fig_fusionb} respectively. The ``$\times$'' fusion involves two loops that are linked to the same base loop, while the ``$\circ$'' fusion involves two loops that carry the same amount of fluxes but are linked to different base loops (Note that these types of fusion were denoted by ``$+$'' and ``$\oplus$'' in Ref.~\onlinecite{threeloop}). One can see that the first four equations resemble Eqs. (\ref{symmetry_2d})-(\ref{linearity_2d_exch}) in 2D systems, while the last two are new to 3D systems. We call (\ref{symmetry3d}) the {\it symmetry relation} and call (\ref{linearity3d1})-(\ref{linearity3d_ex2})  the {\it linearity relations}. Like the 2d relations, the linearity relations encode the fact that braiding and exchanging of loops commute with fusion of loops. The linearity relations (\ref{linearity3d1}, \ref{linearity3d_ex1}) only involve a single base loop and are similar to the 2D linearity relations (\ref{linearity_2d}, \ref{linearity_2d_exch}). The linearity relations (\ref{linearity3d2}, \ref{linearity3d_ex2}) involve the ``$\circ$'' fusion with different base loops, and have no analogue in 2D systems. A graphical proof of (\ref{linearity3d2}) can be found in Ref.\onlinecite{threeloop} and the proof for (\ref{linearity3d_ex2}) is similar.

With the help of these general properties, we will now prove the constraints (\ref{3dconstraint_exch}-\ref{3dconstraint_cyclic2}). The constraints (\ref{3dconstraint_exch}) and (\ref{3dconstraint_sym}) are the simplest to prove, as they follow immediately from (\ref{exch3d}) and (\ref{symmetry3d}). To prove (\ref{3dconstraint_quant}), we first imagine braiding a vortex loop $\alpha$ around $N_j$ identical vortex loops $\beta$, where both $\alpha$ and $\beta$ are linked to a base loop $\gamma$ and $\phi_\alpha= \frac{2\pi}{N_i}e_i$, $\phi_\beta= \frac{2\pi}{N_j}e_j$, $\phi_\gamma= \frac{2\pi}{N_l}e_l$. Clearly the total statistical phase is $N_j\theta_{\alpha\beta,e_l}$. On the other hand, from the linearity property (\ref{linearity3d1}), we know that this phase is equal to that of braiding $\alpha$ around the fusion product of the $\beta$'s. Since the $N_j$ $\beta$'s fuse to a pure charge, it follows that the latter quantity is a multiple of $2\pi/N_i$. Hence, we derive
\begin{equation}
N_j\theta_{\alpha\beta, e_l} =\frac{2\pi}{N_i} \times \text{integer}.
\end{equation}
Comparing with the definition of $\Theta_{ij,l}$, we deduce that
\begin{equation}
N_{ij}\Theta_{ij, l} = 0,
\label{nijeq}
\end{equation}
modulo $2\pi$. Next, we imagine a collection of $N_l$ identical three-loop linked structures. In each structure, loop $\alpha$ and $\beta$ are linked with $\gamma$, and $\alpha$ is braided around $\beta$. The total phase in the $N_l$ braiding processes is $N_l\theta_{\alpha\beta,e_l}$. We then fuse together all the linked structures and make use of the linearity relation (\ref{linearity3d2}), and obtain
\begin{equation}
N_l \theta_{\alpha\beta,e_l} = \theta_{AB, N_le_l}.
\end{equation}
where $A,B$ denote the two loops $A = \alpha\circ\dots\circ\alpha$ and $B =\beta\circ\dots \circ\beta$.
Now since the $N_l e_l = 0$, the two loops $A$ and $B$ are not linked to any base loop so the statistical phase on the right hand side can be computed using the conventional Aharonov-Bohm law (see Ref.~\onlinecite{threeloop} for a more detailed argument):
\begin{equation}
\theta_{AB, N_le_l} =\frac{2\pi}{N_j} q_A \cdot e_j + \frac{2\pi}{N_i} q_B\cdot e_i \label{3dabohm}
\end{equation}
where $q_A, q_B$ denote the amount of charge carried by the (unlinked) loops $A,B$. We conclude that
\begin{equation}
N_l \theta_{\alpha\beta,e_l} = \frac{2\pi}{N_i}\times \text{integer} + \frac{2\pi}{N_j}\times \text{integer}.
\end{equation}
which implies that
\begin{equation}
N_l \Theta_{ij, l} = 0
\label{nleq}
\end{equation}
modulo $2\pi$. Combining (\ref{nijeq}) and (\ref{nleq}), we immediately derive the constraint (\ref{3dconstraint_quant}). The proof of (\ref{3dconstraint_quant_ex}) is similar --- the only difference being that one needs to consider exchange statistics rather than mutual statistics, and the linearity relations (\ref{linearity3d_ex1}, \ref{linearity3d_ex2}) rather than (\ref{linearity3d1}, \ref{linearity3d2}).

The proof of the ``cyclic relations'' (\ref{3dconstraint_cyclic}) and (\ref{3dconstraint_cyclic2}) follows the same philosophy as above, and involves considering certain thought experiments. These thought experiments are described in
Ref.~\onlinecite{threeloop} in the case of $G = (\mathbb Z_N)^K$. It is not hard to extend these thought experiments to $G=\prod_i \mathbb Z_{N_i}$, so we do not repeat them here and instead refer the reader to Ref.~\onlinecite{threeloop}.

It is unfortunate that we are not able to prove the last constraint (\ref{3dconstraint_cyclic3}). Therefore, this relation is just a conjecture. However, from (\ref{3dconstraint_quant_ex}) and (\ref{3dconstraint_cyclic2}), we can prove the weaker constraint $3\Theta_{i,i}=0$.

\subsection{3D general case}

\label{sec:3dnonabelian_constraint}
To complete our discussion, we now consider the case of general 3D gauge theories with gauge group
$G = \prod_{i=1}^K \mathbb Z_{N_i}$. In this case, there are three invariants $\Theta_{i,l}$, $\Theta_{ij,l}$
and $\Theta_{ijk,l}$. We will now argue that $\Theta_{i,l}, \Theta_{ij,l}$
satisfy the constraints (\ref{3dconstraint_exch}-\ref{3dconstraint_quant_ex}), as in the Abelian case. In
addition, the third invariant $\Theta_{ijk,l}$ satisfies
\begin{subequations}
\begin{align}
\Theta_{ijk,l}  & = {\rm sgn}(\hat p) \Theta_{\hat p(i)\hat p(j)\hat p(k), l} \label{3d_antisym_ijk} \\
\Theta_{iij,l}  & = 0 \label{3d_antisym_ijk2} \\
N_{ijk}\Theta_{ijk,l} & = 0 \label{3d_quant_ijk},
\end{align}
\end{subequations}
Unlike the other cases that we have discussed, we expect that the above list of constraints is \emph{incomplete}: that is,
there are likely further constraints on the invariants beyond the ones listed here. One reason for this belief is that in the case of the Dijkgraaf-Witten models, the invariants obey the cyclic constraints (\ref{3dconstraint_cyclic}), (\ref{3dconstraint_cyclic2}), and
(\ref{3dconstraint_cyclic3}), and $\Theta_{ijk,l}$ satisfies the stricter constraint
\begin{align*}
\Theta_{ijk,l}  & = {\rm sgn}(\hat p) \Theta_{\hat p(i)\hat p(j)\hat p(k), \hat p(l)}\\
\Theta_{iij,l}  & = 0 \\
N_{ijkl}\Theta_{ijk,l} & = 0,
\end{align*}
where $\hat p$ is a permutation of the indices $i,j,k,l$ and ${\rm sgn}(\hat p)=\pm$ is its parity. We find it plausible that these
additional constraints may apply more generally than to the Dijkgraaf-Witten models, but we have not been able to prove this fact.

The easiest constraints to establish are (\ref{3d_antisym_ijk}-\ref{3d_quant_ijk}). These constraints follow identical
arguments to the 2D results (\ref{2d_antisym_ijk}-\ref{2d_quant_ijk}). Likewise, the two constraints
(\ref{3dconstraint_exch}) and (\ref{3dconstraint_sym}) follow from the same logic as the 2D results
(\ref{2d_exch}) and (\ref{2d_sym}), which were proved in the general case in section \ref{sec:2dnonabelian_constraint}.

To prove (\ref{3dconstraint_quant}), we first notice that $N_{ij}\Theta_{ij,l}=0$ can be established in the same
way as its 2D analogue (\ref{2d_quant}). Next, we consider a thought experiment with $N_l$ identical three-loop linked structures
$\{\alpha,\beta,\gamma\}$ with $\phi_\alpha=\frac{2\pi}{N_i}e_i, \phi_\beta=\frac{2\pi}{N_j}e_j, \phi_\gamma=\frac{2\pi}{N_k}e_k$. We imagine braiding
each $\alpha$ around the corresponding $\beta$ for $N^{ij}$ times. The result of each braiding process is an
Abelian phase $\Theta_{ij,l}$. Therefore, the result of braiding all the $\alpha$'s simultaneously is
$N_l \Theta_{ij,l}$. Now, since all the phases are Abelian, a linearity relation like (\ref{linearity3d2}) applies
in this case. More specifically, one can argue that the phase associated with this braiding process is
equal to that of braiding the loop $A =\alpha\circ\dots \circ\alpha$ around $B = \beta\circ\dots\circ\beta$ for
$N^{ij}$ times, while both $A,B$ are linked to a base loop which carries a flux $N_l\phi_l$. Now since $N_l \phi_l = 0$ modulo $2\pi$,
$A,B$ are not linked to any base loop so this phase can be computed from the Aharonov-Bohm law, as in equation (\ref{3dabohm}). In this way, we deduce that
\begin{equation}
N_l\Theta_{ij,l} = N^{ij} \left(\frac{2\pi}{N_i}\times\text{integer} + \frac{2\pi}{N_j}\times\text{integer}\right)  = 0.
\end{equation}
Combining this with $N_{ij}\Theta_{ij,l}=0$, we derive the constraint (\ref{3dconstraint_quant}). The proof of
(\ref{3dconstraint_quant_ex}) is similar and involves the use of a linearity relation for $\Theta_{i,l}$ analogous to
(\ref{linearity3d_ex2}).

\section{Do the Dijkgraaf-Witten models exhaust all possible values for the invariants?}
\label{sec:beyondcohomology}

In this section we ask and partially answer the following question:
\begin{description}
\item{\bf Q:} {\it Do there exist Abelian gauge theories for which the invariants acquire values beyond those given by the Dijkgraaf-Witten models? }
\end{description}
If the answer to this question is ``yes'', then it follows that there exist gauge theories that do not belong to the same phase as any of the Dijkgraaf-Witten models. On the other hand, if the answer is ``no'', we cannot make any rigorous statements about the existence or non-existence of gauge theories beyond the Dijkgraaf-Witten models, since we cannot rule out the possibility that two gauge theories may share the same invariants but still belong to distinct phases. That being said, a negative answer can be interpreted as circumstantial evidence that the Dijkgraaf-Witten models exhaust all possible Abelian gauge theories.

To address this question, we compare the general constraints derived in the previous section with our explicit computation of the invariants in the Dijkgraaf-Witten models. We first consider the 2D case. In that case, we know that the invariants must obey constraints (\ref{2d_exch})-(\ref{2d_quant_ex}) as well as (\ref{2d_antisym_ijk})-(\ref{2d_quant_ijk}). At the same time, we know that the Dijkgraaf-Witten models can realize any $(\Theta_i, \Theta_{ij}, \Theta_{ijk})$ of the form given in equations (\ref{theta_value_2d1}) - (\ref{theta_value_2d3}). Comparing these two results, one can easily verify that the Dijkgraaf-Witten models can realize all possible values of the invariants that are consistent with the general constraints. We conclude that in the 2D case, the Dijkgraaf-Witten models exhaust all possible values of the invariants.

Next we consider the 3D Abelian case: that is, 3D gauge theories with gauge group $G$ and Abelian three-loop statistics. In this case, we know that the invariants must obey the constraints (\ref{3dconstraints}). We also know that the Dijkgraaf-Witten models with Abelian statistics\cite{footnote6} can realize any $(\Theta_{i,l}, \Theta_{ij,l})$ of the form given in equations (\ref{theta_value_3d1})-(\ref{theta_value_3d2}). From these two facts, one can show that the Abelian Dijkgraaf-Witten models realize all possible values of the invariants that are consistent with the general constraints; this derivation is given in appendix \ref{sec:appd_formula}. One loophole is that the last constraint (\ref{3dconstraint_cyclic3}) is simply a \emph{conjecture}. Therefore, all we can say is that the Dijkgraaf-Witten models exhaust all possible values of the invariants, assuming this conjecture is correct.

Finally, let us consider the general 3D case: that is, 3D gauge theories with gauge group $G$ and any type of three-loop statistics  (Abelian or non-Abelian). In this case, we have only managed to prove very weak constraints on the invariants, as discussed in the previous section. As a result, the Dijkgraaf-Witten models only realize a small subset of the invariants that are consistent with our constraints. Hence, we cannot make any statements as to whether the Dijkgraaf-Witten models exhaust all possible values of the invariants.

\section{Relation between the invariants and braiding statistics}
\label{sec:relation}

In this section, we show that the topological invariants contain the same information as the full set of braiding statistics data, for 2D or 3D gauge theories with gauge group $G = \prod_{i=1}^K \mathbb Z_{N_i}$ and Abelian statistics. We do not know whether a similar result holds for gauge theories with gauge group $G$ and non-Abelian statistics.
%All we can say is that the invariants contain the full set of information for the case of the Dijkgraaf-Witten models.

\subsection{2D case}

We begin with the 2D case. What we will show is that if two 2D gauge theories with Abelian statistics have the same values for the invariants $\Theta_i$ and $\Theta_{ij}$, then all their braiding statistics must be identical. In this sense the topological invariants contain all the information about the braiding statistics in these systems.

Before presenting our argument, let us recall our definition for when two gauge theories have the ``same'' braiding statistics. As discussed in Sec.~\ref{sec:model_phase}, we say that two gauge theories have the same braiding statistics if there exists a one-to-one correspondence between the quasiparticle excitations in the two theories that (1) preserves all the algebraic structure associated with braiding statistics, e.g. $R$-symbols, fusion rules, $F$-symbols, etc and (2) preserves the gauge flux of excitations. In other words, for each excitation in one gauge theory, there should be a corresponding excitation in the other gauge theory that has the same braiding statistics properties and the same gauge flux.

%Before presenting our argument, we need to explain more precisely what it means for two gauge theories to have identical braiding statistics. Conventionally, two 2D topological phases are said to have the same braiding statistics if there exists a one-to-one correspondence between the quasiparticle excitations in the two theories that preserves all the algebraic structure associated with braiding statistics, e.g. $R$-symbols, fusion rules, $F$-symbols, etc. This notion of equivalence needs to be modified in the case of gauge theories because the excitations in gauge theories have more structure than in general topological phases. In particular, every excitation in a gauge theory can be characterized by the amount of gauge flux that it carries, in addition to its statistical properties. Therefore, in order for two gauge theories to have the same braiding statistics, we not only require there to be a one-to-one correspondence between the excitations in the two theories that preserves all the braiding statistics, but we also require that this correspondence preserves gauge flux: that is, we require that the correspondence maps excitations with a given gauge flux onto other excitations with the same flux.

Given the above definition, our task is as follows. Consider two 2D gauge theories with group $G = \prod_{i}^K \mathbb Z_{N_i}$ and Abelian statistics. Suppose that the gauge theories have the same values for the invariants $\Theta_i$ and $\Theta_{ij}$. We have to show that there exists a one-to-one correspondence between the excitations in the two theories that preserves their exchange statistics, mutual statistics, and gauge flux.

We construct this correspondence as follows. In the first gauge theory, for each $i=1,\dots, K$, we choose one of the $|G|$ types of vortices that carry unit flux $\frac{2\pi}{N_i} e_i$, and denote it by $\hat{v}_i$. Similarly, in the second gauge theory, for each $i=1,\dots, K$, we choose one of the $|G|$ types of vortices carrying unit flux $\frac{2\pi}{N_i} e_i$ and denote it by $\hat{w}_i$. Given that the two gauge theories have identical values of $\Theta_i$ and $\Theta_{ij}$, we know that the exchange statistics and mutual statistics of $\{\hat{v}_i\}$ and $\{\hat{w}_i\}$ are related by
\begin{align}
\theta_{\hat{v}_i} = \theta_{\hat{w}_i} + \frac{2\pi x_i}{N_i}, \quad
\theta_{\hat{v}_i\hat{v}_j} = \theta_{\hat{w}_i\hat{w}_j} + \frac{2\pi y_{ij}}{N^{ij}}  \label{unitflux1}
\end{align}
for some integers $x_i$, $y_{ij}$, with $y_{ii} = 2 x_i$.

In the next step, we fuse some gauge charge $q_i = (q_{i1},...,q_{iK})$ onto each vortex $\hat{w}_i$, to obtain another unit flux vortex $\hat{w}_i'$. We choose the $q_i$ so that the new vortices $\hat{w}_i'$ obey
\begin{align}
\theta_{\hat{v}_i} = \theta_{\hat{w}_i'}, \quad
\theta_{\hat{v}_i\hat{v}_j} = \theta_{\hat{w}_i'\hat{w}_j'} \label{unitflux2}
\end{align}
To see that we can always do this, note that
\begin{align}
\theta_{\hat{w}_i'} &= \theta_{\hat{w}_i} + \frac{2\pi q_{ii}}{N_i}  \nonumber \\
\theta_{\hat{w}_i' \hat{w}_j'} &= \theta_{\hat{w}_i\hat{w}_j} + \frac{2\pi q_{ij}}{N_j} +\frac{2\pi q_{ji}}{N_i}
\end{align}
by the Aharonov-Bohm formula. Hence, we can arrange for equation (\ref{unitflux2}) to hold if we choose the gauge charges $q_i$ so that
they satisfy
\begin{eqnarray}
q_{ii} &=& x_i \ \ \ (\text{mod } N_i) \nonumber \\
\frac{1}{N_{ij}}(N_i q_{ij} + N_j q_{ji} )&=& y_{ij} \ \ \ (\text{mod } N^{ij})
\end{eqnarray}

We are now ready to construct the desired one-to-one correspondence between the excitations in the two gauge theories. We note that every excitation in the first gauge theory can be written uniquely as a fusion product
\begin{equation}
\alpha = (\hat{v}_1)^{a_1} \times \cdots \times (\hat{v}_K)^{a_K} \times q.
\label{rep1}
\end{equation}
where $a_i$ are integers with $0 \leq a_i \leq N_i-1$, and where $q = (q_1,\dots, q_K)$ is some gauge charge with $0 \leq q_i \leq N_i-1$. Similarly,
every excitation in the second gauge theory can be written uniquely as
\begin{equation}
\alpha' = (\hat{w}_1')^{a_1} \times \cdots \times (\hat{w}_K')^{a_K} \times q.
\label{rep2}
\end{equation}
We define a one-to-one correspondence between two sets of excitations by mapping
\begin{eqnarray}
\alpha &=& (\hat{v}_1)^{a_1} \times \cdots \times (\hat{v}_K)^{a_K} \times q  \nonumber \\
\leftrightarrow \alpha' &=& (\hat{w}_1')^{a_1} \times \cdots \times (\hat{w}_K')^{a_K} \times q
\label{2dmap}
\end{eqnarray}
It is clear that this correspondence preserves gauge flux since
\begin{equation}
\phi_\alpha = \left(\frac{2\pi a_1}{N_1},\dots, \frac{2\pi a_k}{N_k} \right) = \phi_{\alpha'}
\end{equation}
To see that this correspondence preserves the exchange statistics and mutual statistics of the excitations, we need to check that
\begin{align}
\theta_{\alpha} = \theta_{\alpha'}, \quad \theta_{\alpha \beta} = \theta_{\alpha' \beta'}
\end{align}
for any $\alpha, \beta$ in the first gauge theory and corresponding $\alpha', \beta'$ in the second gauge theory. These relations follow immediately from (\ref{unitflux2}) together with the linearity relations (\ref{linearity_2d}) and (\ref{linearity_2d_exch}). This completes our argument: we have shown that if two gauge theories have the same values of the invariants $\Theta_i, \Theta_{ij}$, then all their braiding statistics is identical.

\subsection{3D case}

%As before, we can attach charge to a vortex loop. Similar to the 2D case, there is an issue of {\it absolute} charge versus {\it relative} charge on a vortex loop. A complication in the 3D case is that loops can come as being \emph{linked} or \emph{unlinked} to other loops. The two types behave differently on this issue. For unlinked loops\cite{footnote2}, we do have a well-defined notion of absolute charge on them. This is because we can shrink an unlinked loop to a point and measure the charge of this point. In particular, there exist {\it neutral} loops which can be shrunk to a point and annihilated with nothing left. On the other hand, if a vortex loop $\alpha$ is linked with another loop $\gamma$, we can only define a {\it relative} charge of $\alpha$ with respective to a {\it reference } loop $\hat \alpha$ which is also linked to $\gamma$.  The reason is similar to that for 2D vortices: $\alpha$ is either created a vortex-antivortex pair $\alpha$ and $\bar \alpha$, or created alone but at some point of the creation process it touches the loop $\gamma$. That is, the creation process necessarily involves two loops, either $\alpha$ and $\bar\alpha$, or $\alpha$ and $\gamma$. Therefore, it is not possible to assign a universal {\it absolute} charge to the loop $\alpha$ alone. Details about this issue are discussed in Sec.~\ref{sec:absolutecharge}. In any way, we will see that the notion of relative charge is enough for us to study loop braiding statistics.

We now repeat the argument in the 3D case. Consider two 3D gauge theories with group $G = \prod_{i}^K \mathbb Z_{N_i}$ and Abelian three-loop statistics. Suppose that the gauge theories have the same values for the invariants $\Theta_{i,l}$ and $\Theta_{ij,l}$. We will show that the two gauge theories have identical three-loop statistics. More precisely, we will show that for each gauge flux $\phi$, there exists a one-to-one correspondence between the loop-like excitations in the two theories
that are linked with base loops with flux $\phi$, such that the corresponding excitations have the same three-loop statistics and the same gauge flux.

The derivation closely follows the 2D case. To begin, we focus on the first gauge theory, and we fix a base loop that carries unit flux $\frac{2\pi}{N_l} e_l$. For each $i=1,...,K$, we choose one of the $|G|$ types of vortex loops that carry flux $\frac{2\pi}{N_i} e_i$ and are linked with the base loop and we denote it by $\hat{v}_i$. We repeat this process for the second gauge theory. That is, we fix a base loop with flux $\frac{2\pi}{N_l} e_l$ and we choose vortex loops $\{\hat{w}_i\}$ that are linked with the base loop and carry flux $\frac{2\pi}{N_i} e_i$. Using the fact that the two gauge theories have identical values of $\Theta_{i,l}$ and $\Theta_{ij,l}$,
we know that the three-loop statistics of the $\{\hat{v}_i\}$ and $\{\hat{w}_i\}$ loops are related by
\begin{align*}
\theta_{\hat{v}_i,e_l} = \theta_{\hat{w}_i,e_l} + \frac{2\pi x_{i,l}}{N_i}, \quad
\theta_{\hat{v}_i\hat{v}_j,e_l} = \theta_{\hat{w}_i\hat{w}_j,e_l} + \frac{2\pi y_{ij,l}}{N^{ij}}
\end{align*}
for some integers $x_{i,l}$, $y_{ij,l}$, with $y_{ii,l} = 2 x_{i,l}$. Note that the integers $x_{i,l}$ and $y_{ij,l}$ may depend on $l$ --- that is they may take different values for different base loops.

We next fuse some gauge charge $q_i = (q_{i1},...,q_{iK})$ onto each vortex loop $\hat{w}_i$, to obtain another vortex loop $\hat{w}_i'$. We choose the $q_i$ so that the new vortex loops $\hat{w}_i'$ obey
\begin{align}
\theta_{\hat{v}_i,e_l} = \theta_{\hat{w}_i',e_l}, \quad
\theta_{\hat{v}_i\hat{v}_j,e_l} = \theta_{\hat{w}_i'\hat{w}_j,e_l} \label{3dunitflux2}
\end{align}
The fact that we can always find such a $q_i$ follows from the same reasoning as in the 2D case.

We now construct a one-to-one correspondence between the loop excitations in the two gauge theories, focusing on the excitations that are linked with a base loop with flux $\frac{2\pi}{N_l} e_l$. The construction is identical to the 2D case. First, we note that every loop excitation $\alpha$ in the first gauge theory can be written uniquely as a fusion product of the $\hat{v}_i$ vortex loops together with some gauge charge, as in equation (\ref{rep1}). Similarly, every loop excitation in the
second gauge theory can be written uniquely as a product of the $\hat{w}_i'$ vortex loops as in equation (\ref{rep2}). We can therefore define a one-to-one correspondence using the mapping in equation (\ref{2dmap}). For the same reasons as in the 2D case, it is clear that this correspondence preserves the statistics of the loop excitations, as well as their gauge flux.

To finish the derivation, we need to generalize the above one-to-one correspondence to the case where the base loop carries arbitrary flux $\phi$. This generalization is easy to prove since we can construct base loops with arbitrary flux by fusing together base loops with unit flux using the ``$\circ$'' fusion process (see Fig.~\ref{fig_fusionb}). Furthermore, using the linearity relations (\ref{linearity3d2}, \ref{linearity3d_ex2}) we can see that the three-loop statistics associated with these more general base loops is completely determined by the three-loop statistics for the unit flux base loops. Putting these pieces together, we can construct a similar one-to-one correspondence between loop excitations that are linked with base loops with arbitrary flux. This completes our argument and proves that the two gauge theories have identical three-loop statistics.

%\subsection{Absence of non-Abelian statistics in $\mathbb Z_N$ and $\mathbb Z_{N_1}\times \mathbb Z_{N_2}$ gauge theories}

\section{Conclusion}
\label{sec:conclusion}

In this paper, we have studied the braiding statistics of 2D and 3D gauge theories with group $G = \prod_{i=1}^K \mathbb Z_{N_i}$, and we have defined topological invariants that summarize some of the most important aspects of this braiding structure. In the 2D case, these invariants consist of three tensors, $\{\Theta_i, \Theta_{ij},\Theta_{ijk}\}$, while in the 3D case,  they consist of three tensors, $\{\Theta_{i,l}, \Theta_{ij,l},\Theta_{ijk,l}\}$. These tensors are defined in terms of certain composite braiding processes involving vortices and vortex loops.

Using these invariants, we have obtained two results. First, we have shown that the invariants distinguish all 2D and 3D Dijkgraaf-Witten models ($=$ gauged group cohomology models) with group $G$. Second, we have shown that the Dijkgraaf-Witten models with group $G$ exhaust all possible values of the invariants in the 2D case and we have derived similar, but weaker, results in the 3D case.

So far our discussion has focused on gauge theories, or more precisely, gauged SPT models. We now return to the questions raised in the introduction, and discuss the implications of our findings for \emph{ungauged} SPT models. We begin with our result that the invariants take different values for each of the 2D and 3D Dijkgraaf-Witten models ($=$ gauged group cohomology models) with group $G$. This result has an immediate implication: the 2D and 3D group cohomology models with group $G$ all belong to distinct phases.

Next, we consider our finding that the 2D Dijkgraaf-Witten models exhaust all possible values for the invariants. While we cannot draw rigorous conclusions from this result, it is at least consistent with the possibility that the 2D group cohomology models realize all possible SPT phases with symmetry group $G$. Our 3D results on this topic are also consistent with this possibility, though they are somewhat weaker.

In short, our results support the group cohomology classification conjecture of Chen, Gu, Liu, and Wen\cite{chen13}, for the case of finite, Abelian symmetry group $G$. In addition, our results allow us to go further: they provide a simple diagnostic for determining whether a specific microscopic Hamiltonian belongs to the same phase as a specific group cohomology model. The diagnostic involves gauging the Hamiltonian and then computing the topological invariants of the associated gauge theory. If the invariants for the microscopic Hamiltonian are different from that of the group cohomology model, then we may conclude that the Hamiltonian belongs to a distinct phase. If the invariants are the same, then it may belong to the same phase, though we cannot be certain because we do not know if the invariants are complete in the sense that they distinguish all possible SPT phases.

%An obvious question for future work is to determine whether the invariants are complete in the above sense (this is related to question $3$ from the introduction). Another possible direction is to study braiding statistics in gauge theories with non-Abelian gauge group and see if similar topological invariants can be defined in that context. Finally, it would be interesting to study gauge theories associated with fermionic SPT models. Such systems may have an even richer braiding statistics structure then the bosonic case, analyzed here.

We see this work as a first step towards answering the three questions raised in the introduction. We have managed to make some progress on these questions for the special case of Abelian symmetry groups, but many issues remain unresolved. First, although we have partially answered questions 1-2 from the introduction, we have said nothing at all about question 3 --- which asks whether the braiding statistics data can distinguish all possible SPT phases. Another important direction for future work is to understand the braiding statistics in gauge theories with \emph{non-Abelian} gauge group and see if similar topological invariants could be defined in that context. Finally, it would be interesting to study gauge theories associated with fermionic SPT models. Such systems may have an even richer braiding statistics structure than the bosonic case analyzed here.

\begin{acknowledgments}
We thank C.-H. Lin, Z. Wang, J. Wang, X. Chen and P. Ye for helpful discussions. This work is supported by the Alfred P. Sloan foundation and NSF under grant No. DMR-1254741.
\end{acknowledgments}

\appendix

\section{Gauging prescription}
\label{sec:appd_gauging}

In this Appendix, we give a general prescription for how to gauge a lattice boson model with an Abelian symmetry group $G = \prod_{i=1}^K\mathbb Z_{N_i}$. The procedure we describe follows the usual minimal coupling scheme for lattice gauge theories\cite{kogut79}. The only nonstandard element is that we will set the gauge coupling constant to zero in order to
maximize our control over the models, as in Refs.~\onlinecite{levin12,gu14}. More precisely, what we mean by this is that the Hamiltonians for the gauged models commute with the flux operators that measure the gauge flux through each plaquette in the lattice. This property has a nice consequence: the gauge theories we construct are guaranteed to be gapped and deconfined as long as the original boson models are gapped and don't break the symmetry spontaneously.

For concreteness, we will focus on a particular kind of boson model with $\prod_{i=1}^K\mathbb Z_{N_i}$ symmetry. Specifically, we will focus on lattice boson models built out of $K$ species of bosons, where the particle number of the $i$th species is conserved modulo $N_i$ and where the different species of bosons live on the sites $p$ of some 2D or 3D lattice. We denote the boson creation operator for the $i$th species on lattice site $p$ by $b_{p,i}^\dagger$. We assume that the boson model has local interactions, so that the Hamiltonian of the bosons can be written as
\begin{equation}
H= \sum_A H_A(\{b_{p,i}\}) \label{hamiltonian}
\end{equation}
where the sum is taken over localized regions $A$, and where $H_A$ is some operator composed out of boson creation and annihilation operators acting on region $A$.

We now discuss our procedure for gauging such a Hamiltonian. The first step is to introduce a Hilbert space $\mathcal H_{pq}$ of dimension $|G|=\prod_{i=1}^K N_i$ for each link $pq$ of the lattice. This Hilbert space is spanned by basis states $\{ |m\rangle \}$, where $m=(m_1,\dots, m_K)$ with $0\le m_i\le N_i-1$. Along with the Hilbert space, we introduce lattice gauge field operators $\mu_{pq,i}$ for each species $i=1,\dots, K$. These operators are defined by
\begin{equation}
\mu_{pq,i} |m\rangle = e^{\pm i\frac{2\pi}{N_i} m_i} |m\rangle.
\end{equation}
%Here, we let each link be associated with a direction, so that we can defined $u_{qp} \equiv u_{pq}^\dag$.
with a $+$ or $-$ sign depending on whether $pq$ is parallel or anti-parallel to some fixed orientation that we assign to every link of the lattice. Likewise, we define a set of unitary shift operators $S_{pq,a}$ for each group element $a\in G$:
\begin{equation}
S_{pq,a} |m\rangle = |m\pm a\rangle.
\end{equation}
with a $+$ or $-$ sign depending on whether $pq$ is parallel or antiparallel to the prescribed orientation. Here, we have used $a=(a_1,\dots, a_k)$, $0\le a_i \le N_i-1$, to denote the group elements of $G = \prod_{i=1}^K\mathbb Z_{N_i}$.

In the second step, we replace each operator $H_A(\{b_{p,i}\})$ by a gauged operator $\tilde H_A(\{b_{p,i}, \mu_{pq,i}\})$ following the minimal coupling procedure:
\begin{equation}
H_A(\{b_{p,i}\}) \rightarrow \tilde H_A(\{b_{p,i}, \mu_{pq,i}\}),
\end{equation}
For example, a nearest neighbor hopping term undergoes the following substitution under minimal coupling
\begin{align}
b_{p,i} b_{q,i}^\dag & \rightarrow b_{p,i} b_{q,i}^\dag \mu_{pq,i}
\end{align}
%A more complicated example is
%\begin{align}
%(b_{pi} b_{qi}^\dag)^s (b_{qj} b_{pj}^\dag)^t & \rightarrow (b_{pi} b_{qi}^\dag)^s (b_{qj} b_{pj}^\dag)^t (\mu_{pq,i})^s(\mu_{pq,j}^\dag)^t.
%\end{align}

For more complicated terms involving multiple sites, the substitution contains a product of $\mu$ operators acting on a path that connects the sites. One subtlety is that there is an ambiguity for how to choose the path. This ambiguity is eliminated by the third step in the gauging procedure. In this step, we multiply $\tilde H_A$ by a projection operator $P_A$ which projects onto the states that have vanishing gauge flux through each plaquette that belong to $A$. That is, the projector $P_A$ can be written as a product
\begin{equation}
P_A = \prod_{\<pqr\> \in A} P_{\<pqr\>}
\end{equation}
where $P_{\langle pqr \rangle}$ projects onto states with vanishing flux through a particular plaquette $\langle pqr \rangle$, which we have assumed to be triangular for concreteness. The projector $P_{\langle pqr\rangle}$ can be explicitly written as
\begin{equation}
P_{\<pqr\>} =\frac{1}{|G|}\prod_{i=1}^K \left( \sum_{k=0}^{N_i-1}(\mu_{pq,i}\mu_{qr,i}\mu_{rp,i} )^k\right). \label{projector}
\end{equation}
After multiplying $\tilde H_A$ by $P_A$, one can show that all paths enclosed by $A$ lead to the same term so that the minimal coupling procedure is unambiguous. It is not hard to see that $P_{\<pqr\>}$ is a Hermitian operator.

In the last step of the gauging procedure, we add a term of the form $-\Delta P_{\<pqr \>}$ to the Hamiltonian for each plaquette, so that it costs finite energy to create vortex excitations. After applying these steps, the final gauged Hamiltonian is
\begin{equation}
\tilde H = \sum_A  \tilde H_A(\{b_{p,i}, \mu_{pq,i}\}) P_A  - \Delta \sum_{\<pqr \>} P_{\<pqr \>}
\end{equation}
where we assume $\Delta$ is large and positive. This Hamiltonian is defined on a Hilbert space consisting of gauge invariant states, i.e. states $|\Psi\rangle$
satisfying
\begin{equation}
T_{p,a}|\Psi\rangle= |\Psi\rangle
\end{equation}
for every $p,a$, where $T_{p,a}$ is the gauge transformation associated with group element $a \in G$ and site $p$:
\begin{equation}
T_{p,a} = e^{i\sum_i \frac{2\pi a_i}{N_i} b_{p,i}^\dag b_{p,i}} \prod_{q \in \text{neigh.($p$)} } S_{qp,a}
\end{equation}
This constraint is the analogue of the usual Gauss's law of electromagnetism, $\nabla\cdot E = \rho$.

The gauging prescription we have just described has a special property: the Hamiltonian $\tilde{H}$ commutes
with the flux through each plaquette of the lattice. That is, $[\tilde{H}, \mu_{pq,i}\mu_{qr,i}\mu_{rp,i}] = 0$ for every plaquette
$\<pqr\>$ and every $i=1,...,K$. This property has an important consequence: the gauge theory $\tilde{H}$ is guaranteed to be gapped and deconfined
as long as $H$ is gapped and doesn't break the symmetry
spontaneously. To see this, note that $[\tilde{H},P_{\<pqr\>}] = 0$, so the eigenstates of $\tilde H$ are also eigenstates of
$P_{\<pqr\>}$. As long as $\Delta$ is large, then all the low energy eigenstates  $|\Psi\>$ will have vanishing flux: that is,
$P_{\<pqr\>} |\Psi\> = |\Psi\>$. At the same time, it is easy to see that, within the zero flux sector, the energy spectrum of
$\tilde{H}$ is identical to the energy spectrum of the original boson model $H$. We conclude that $\tilde{H}$ and $H$ have identical
low energy spectra. Hence, $\tilde{H}$ is guaranteed to be gapped if $H$ is gapped. Similar reasoning shows that $\tilde{H}$ is
deconfined as long as $H$ doesn't break the symmetry spontaneously.

%The gauged system $\tilde H$ may belong to different phases depending on the original system $H$.  Here, we are interested in the topological phases of gauge theories for a fixed $G$: if two gauge theories $\tilde H_1$ and $\tilde H_2$  can be smoothly connected without closing the energy gap, they belong to the same phase, otherwise they belong to distinct phases. One of our tasks in this paper is to use braiding statistics to distinguish various phases of gauge theories.

%The state The global $\prod_{i=1}^K\mathbb Z_{N_i}$ symmetry is then promoted to a local $\prod_{i=1}^K\mathbb Z_{N_i}$ gauge symmetry. Nontrivial braiding statistics will emerge in this {\it gauged} model, and the type of statistics depends on, and by conjecture\cite{levin12} identifies, the topological properties of the original {\it ungauged} model.

\section{Group cohomology}
\label{sec:appd_cohomology}

In this Appendix, we review the basic ingredients of the cohomology of finite groups.\cite{chen13, propitius95, brown} We focus on the cohomology group $H^n[G,U(1)]$.

Let $G$ be a finite group. The basic objects that group cohomology studies are \emph{$n$-cochains}. An $n$-cochain is a $U(1)$ valued function $c(g_1, \dots, g_n)$:
\begin{displaymath}
c: \underbrace{G\times G\times \cdots \times G}_{n \text{ times}} \rightarrow U(1).
\end{displaymath}
The collection of $n$-cochains form an Abelian group $\mathcal C^n$, where the group operation is defined by
\begin{displaymath}
(c_1\cdot c_2)(g_1, \dots, g_n) = c_1(g_1, \dots, g_n)\cdot c_2(g_1, \dots, g_n).
\end{displaymath}
The \emph{coboundary operator} $\delta$ is a map $\delta: \mathcal C^n\rightarrow \mathcal C^{n+1}$, defined by
\begin{align}
\delta c &(g_1, \dots, g_{n+1}) = c(g_2, \dots, g_{n+1})c(g_1, \dots, g_n)^{(-1)^{n+1}} \nonumber\\
& \times \prod_{i=1}^n [c(g_1, \dots, g_i g_{i+1}, \dots, g_{n+1})]^{(-1)^{i}}. \label{delta}
\end{align}
It is easy to check that the coboundary operator satisfies $\delta(c_1\cdot c_2) = \delta c_1 \cdot \delta c_2$. More importantly, one can check that
$\delta$ is nilpotent: $\delta^2=1$.

With the help of the coboundary operator, we can now define \emph{$n$-cocycles} and \emph{$n$-coboundaries}. An $n$-cocycle is an $n$-cochain $\omega$ that
satisfies $\delta \omega =1$. For example, 3-cocycles satisfy
\begin{equation}
\frac{\omega(g_2,g_3,g_4)\omega(g_1,g_2g_3,g_4)\omega(g_1,g_2,g_3)}{\omega(g_1g_2,g_3,g_4)\omega(g_1,g_2,g_3g_4)}=1, \label{3cocyclecond}
\end{equation}
and 4-cocycles satisfy
\begin{equation}
\frac{\omega(g_2,g_3,g_4,g_5)\omega(g_1,g_2g_3,g_4,g_5)\omega(g_1,g_2,g_3,g_4g_5)}{\omega(g_1g_2,g_3,g_4,g_5)\omega(g_1,g_2,g_3g_4,g_5)\omega(g_1,g_2,g_3,g_4)}=1.
\label{4cocyclecond}
\end{equation}
Likewise, an $n$-coboundary is an $n$-cochain $\nu$ that can be written as $\nu =\delta  c$ where $c\in\mathcal C^{n-1}$.
The nilpotence of $\delta$ implies that a coboundary must also be a cocycle. This allows us to define an equivalence relation for the cocycles: two $n$-cocycles
$\omega_1$ and $\omega_2$ are said to be {\it cohomologically equivalent} if and only if $\omega_1 = \omega_2 \cdot \delta c$, for some $c\in \mathcal C^{n-1}$.
The equivalence classes of the $n$-cocycles form an Abelian group, called the $n$th \emph{cohomology group}, which is denoted by $H^{n}[G, U(1)]$.

In this paper, we will only need the cohomology group $H^{n}[G, U(1)]$ for $G= \prod_{i=1}^K \mathbb Z_{N_i}$ and for $n=3$ and $4$. These cohomology groups can be computed explicitly using the Kunneth formula:\cite{propitius95, chen13}
\begin{align}
H^3[G, U(1)]  & = \prod_{ i} \mathbb Z_{N_i}\prod_{i < j} \mathbb Z_{N_{ij}}\prod_{i < j <k} \mathbb Z_{N_{ijk}} \\
H^4[G, U(1)]  & =\prod_{i < j} \left(\mathbb Z_{N_{ij}}\right)^2\prod_{ i < j <k} \left(\mathbb Z_{N_{ijk}}\right)^2 \prod_{ i < j <k<l} \mathbb Z_{N_{ijkl}}.
\end{align}

\section{Group cohomology models and Dijkgraaf-Witten models}
\label{sec:appd_dijkgraafwitten}

In this Appendix, we briefly review the group cohomology models of Ref.~\onlinecite{chen13}, as well as the Dijkgraaf-Witten models of Ref.~\onlinecite{dijkgraaf90}. In addition, we show that coupling the group cohomology models to a lattice gauge field gives exactly the Dijkgraaf-Witten models. For convenience, we describe these models as well as the gauging procedure using a path integral formulation in Euclidean space-time. This is different from the gauging procedure in Sec.~\ref{sec:appd_gauging} which is described in a Hamiltonian formulation. We expect the similar results can be derived in a Hamiltonian formulation (e.g., see Ref.~\onlinecite{hu13,wan14} for a Hamiltonian description of Dijkgraaf-Witten models).

\subsection{Group cohomology models}
The basic data needed to construct a $d+1$-dimensional group cohomology model with (finite) group $G$ is (1) a $d+1$-cocycle $\omega$ together with (2) a triangulation of $d+1$-dimensional Euclidean space-time. To build the model, we label the vertices of the triangulation by an ordered sequence $i, j, \dots$, the links by $[ij], [jk], \dots$, and the triangular plaquettes by $[ijk], \dots$, etc. We will refer to the vertices as ``0-simplices'', the links as ``1-simplices'' the triangular plaquettes as ``2-simplices'' and so on. The basic degrees of freedom in the model are group elements $g_i \in G$ that live on the vertices $i$ of the triangulation. For every space-time
configuration $\{g_i\}$, we assign a local weight $[\omega(g_{i}^{-1}g_{j}, \dots, g_{k}^{-1}g_{l})]^{\sigma_{ij\dots kl}}$ to each $d+1$-simplex $[ij\dots kl]$ ($i<j<\dots<k<l$) where $\sigma_{ij\dots kl}=\pm 1$ is the chirality of the simplex $[ij\dots kl]$. The action corresponding to $\{g_i\}$ is given by the product of the local weights
\begin{equation}
e^{-S_1(\{g_{i}\})} =  \prod_{[ij\dots kl]}  [\omega(g_{i}^{-1}g_{j}, \dots, g_{k}^{-1}g_{l})]^{\sigma_{ij\dots kl}}.  \label{spt_action}
\end{equation}
Summing over all the configurations $\{g_i\}$, we obtain the partition function
\begin{equation}
Z_1 = \frac{1}{|G|^{N_v}}\sum_{\{g_i\}} e^{-S_1(\{g_{i}\})}, \label{spt_partition_fuction}
\end{equation}
where $\frac{1}{|G|^{N_v}}$ is a normalization factor, $|G|$ is the size of the group, and $N_v$ is the number of vertices. One can easily check that the action (\ref{spt_action}) is invariant under the global symmetry
\begin{equation}
g_i\rightarrow gg_i,
\end{equation}
for all $g\in G$. According to the arguments in Ref.~\onlinecite{chen13}, the ground states of these models are gapped and short-range entangled for any $G$ and $\omega$. Moreover, two cocycles that differ by a coboundary define the same group cohomology model. Thus, the group cohomology models are labeled by equivalence classes of cocycles, i.e., by elements of the cohomology group $H^{d+1}[G, U(1)]$.

 \subsection{Dijkgraaf-Witten models}

The basic data needed to construct a $d+1$-dimensional Dijkgraaf-Witten model with (finite) group $G$ is the same as that for a group cohomology model: (1) a $d+1$-cocycle $\omega$ together with (2) a triangulation of $d+1$-dimensional Euclidean space-time. Unlike the group cohomology SPT models, the basic degrees of freedom in a Dijkgraaf-Witten model are group elements $h_{ij} \in G$ that live on the \emph{links} $[ij]$ of the triangulation. For every space-time
configuration $\{h_{ij}\}$, the corresponding action $e^{-S_2(\{h_{ij}\})} $ is defined as follows. First, one needs to determine if the configuration is flat, i.e., $h_{ij}h_{jk}h_{ki}=1$ for every 2-simplex $[ijk]$. If it is flat, we assign a local weight $[\omega(h_{ij}, \dots, h_{kl})]^{\sigma_{ij\dots kl}}$ to each $d+1$-simplex $[ij\dots kl]$ ($i<j<\dots<k<l$) where $\sigma_{ij\dots kl}=\pm 1$ is the chirality of the simplex $[ij\dots kl]$. The action is then given by
\begin{equation}
e^{-S_2(\{h_{ij}\})} =  \prod_{[ij\dots kl]}  [\omega(h_{ij}, \dots, h_{kl})]^{\sigma_{ij\dots kl}}.
\end{equation}
If the gauge configuration is not flat, then $e^{-S_2(\{\{h_{ij}\})}=0$. Summing over all the configurations $\{h_{ij}\}$, we obtain the partition function
\begin{equation}
Z_2 = \frac{1}{|G|^{N_v}}\sideset{}{'}\sum_{\{h_{ij}\}}  \prod_{[ij\dots kl]}  [\omega(h_{ij}, \dots, h_{kl})]^{\sigma_{ij\dots kl}}, \label{dw_partition_fuction}
\end{equation}
where $\frac{1}{|G|^{N_v}}$ is again a normalization factor, and $\sum'$ is a summation over flat gauge configurations. The partition function $Z_2$ describes the Dijkgraaf-Witten models. One can check that if two cocycles $\omega, \omega'$ differ by a coboundary, then they define the same Dijkgraaf-Witten models. Thus, the Dijkgraaf-Witten models are labeled by elements of $H^{d+1}[G,U(1)]$, just like the group cohomology models.

\subsection{Connection between the two classes of models}
We now show that the Dijkgraaf-Witten model with group $G$ and cocycle $\omega$ is equivalent to the group cohomology model of the same group $G$ and cocycle $\omega$ after the global symmetry of the latter is gauged.

To gauge the symmetry in the group cohomology model, we introduce lattice gauge fields $h_{ij}\in G$ that live on the links $[ij]$ of the triangulation. We then couple the matter degrees of freedom $\{g_i\}$ and gauge degrees of freedom $\{h_{ij}\}$ by replacing each $g_i^{-1} g_j$ in the action (\ref{spt_action}) by $ g_i^{-1} h_{ij} g_j$, following the minimal coupling procedure. After this step, the action becomes
\begin{equation}
e^{-\tilde S_1(\{g_i\}, \{h_{ij}\})} =  \prod_{[ij\dots kl]} \left[\omega(g^{-1}_i h_{ij} g_j, \dots, g^{-1}_k h_{kl} g_l)\right]^{\sigma_{ij\dots kl}}.
\end{equation}
The next step is to choose a value for the gauge coupling constant. Here, as in Appendix \ref{sec:appd_gauging}, we choose the gauge coupling constant to be $0$. This means that we set $e^{-\tilde S_1(\{g_i\}, \{h_{ij}\})}=0$ if the gauge configuration is not flat. With this choice of coupling constant, the gauged partition function acquires the form
\begin{align}
\tilde Z_1  = &\frac{1}{|G|^{2N_v}}\sum_{\{g_i\}} \sideset{}{'}\sum_{\{h_{ij}\}} \nonumber\\
& \prod_{[ij\dots kl]} \left[\omega(g^{-1}_i h_{ij} g_j, \dots, g^{-1}_k h_{kl} g_l)\right]^{\sigma_{ij\dots kl}},
\end{align}
where the summation $\sum'$ is taken only over flat gauge configurations, and where we have included a normalization factor $\frac{1}{|G|^{N_v}}$. By construction, the partition function $\tilde Z_1$ has a local gauge symmetry
\begin{equation}
g_i\rightarrow \alpha_i g_i, \quad h_{ij}\rightarrow \alpha_i h_{ij}\alpha_j^{-1}, \label{gaugetransformation}
\end{equation}
for all $\alpha_i\in G $ and all $i,j.$

To see the connection between the gauged group cohomology models $\tilde Z_1$  and the Dijkgraaf-Witten models $Z_2$,  we fix the gauge in  $\tilde Z_1$. The gauge that we choose is
\begin{equation}
g_i=1,  \quad \text{for all } i.
\end{equation}
In this gauge, the partition function $\tilde Z_1$ becomes
\begin{equation}
\tilde Z_1  = \frac{1}{|G|^{N_v}}\sideset{}{'}\sum_{\{h_{ij}\}}\prod_{[ij\dots kl]} \left[\omega(h_{ij}, \dots, h_{kl} )\right]^{\sigma_{ij\dots kl}},
\end{equation}
where the numerical factor $|G|^{N_v}$ comes from performing the sum over $\{g_i\}$. We can see that $\tilde Z_1$ is identical to $Z_2$, proving that the gauged group cohomology model with group $G$ and cocycle $\omega$ is equivalent to the Dijkgraaf-Witten model with the same $G$ and $\omega$.

\section{Properties of fusion rules in Abelian discrete gauge theories}

\label{sec:appd_fusionprop}
In this Appendix, we derive some properties of the fusion rules of excitations in 2D gauge theories with group $G = \prod_i \mathbb Z_{N_i}$.

The first property is that any excitation $\gamma$ that appears in the fusion product $\alpha\times\beta = \sum_\gamma N_{\alpha\beta}^\gamma \gamma$ must obey $\phi_\gamma = \phi_\alpha+\phi_\beta$. This property is clear from the following thought experiment. Imagine braiding an arbitrary charge $q$ around $\alpha$ and $\beta$. One can braid $q$ around $\alpha$ and $\beta$ sequentially, which gives a phase $q\cdot(\phi_\alpha+\phi_\beta)$. Or, one can first fuse $\alpha$ and $\beta$ into some $\gamma$, then braid $q$ around $\gamma$, leading to a phase $q\cdot\phi_\gamma$. Clearly the two processes should give the same phase, so $q\cdot(\phi_\alpha+\phi_\beta) = q\cdot \phi_\gamma$. Since $q$ is arbitrary, we have $\phi_\alpha+\phi_\beta=\phi_\gamma$.

Another property of the fusion rules is that when an excitation $\alpha$ is fused with a charge $q$, there is exactly one fusion outcome:
\begin{equation}
q\times \alpha  = \alpha', \label{qa}
\end{equation}
where $\alpha'$ might be the same as $\alpha$. To prove this property, we imagine four excitations $\alpha, \bar \alpha$ and $q, \bar q$, where $\bar{\alpha}$ denotes the antiparticle of $\alpha$ and we suppose that the overall fusion channel for the four excitations is the vacuum. We now count the degeneracy of this four-excitation space in two different ways. First, we note that $q, \bar q$ are Abelian particles, so they must fuse to the vacuum, which in turn forces $\alpha, \bar\alpha$ to fuse to the vacuum. So, the four-excitation space is non-degenerate. On the other hand, we can also fuse the particles in a different order: we first fuse $q$ with $\alpha$ and fuse $\bar q$ with $\bar \alpha$, then fuse the resulting particles. If we fuse the particles in this way, we can see that the degeneracy is given by the number of different fusion outcomes in $q \times \alpha$. We conclude that there is a unique fusion outcome in $q \times \alpha$. Therefore, (\ref{qa}) holds.

%Imagine four excitations $\alpha, \bar \alpha$ and $q, \bar q$, where $\bar{\alpha}$ denotes the antiparticle of $\alpha$. Suppose that the overall fusion channel for the four excitations is the vacuum. We now count the degeneracy of this four-excitation space in two different ways. First, we note that $q, \bar q$ are Abelian particles, so they must fuse to the vacuum, which in turn forces $\alpha, \bar\alpha$ to fuse to the vacuum. So, the four-excitation space is non-degenerate. Now suppose we have a fusion rule of the form $q\times \alpha  = \alpha'+\alpha''$, and consider the above four-excitation state again. This time, we first fuse $q$ with $\alpha$ and fuse $\bar q$ with $\bar \alpha$, then fuse the resulting particles. Obviously, we find that the degeneracy is $2$, because $q$ can be fused with $\alpha$ in two ways. This results in a contradiction. This contradiction also eliminates the special case where $\alpha'=\alpha''$, i.e., where the coefficient of $\alpha'$ is $2$. Therefore, (\ref{qa}) holds.

The third property is that an excitation $\alpha$ and its anti-particle $\bar \alpha$ can only fuse to charges, i.e.
\begin{equation}
\alpha\times \bar \alpha = \emptyset + q_1 + q_2 +\dots. \label{E1}
\end{equation}
where $\emptyset$ denotes the vacuum. Moreover, the coefficient of each $q_i$ appearing in the fusion rule is 1. The first statement is easy to prove: all particles on the right side must be charges since $\phi_\alpha+\phi_{\bar\alpha}=0$. To prove that the coefficient associated with each $q_i$ is 1, we use the fact that $N_{\alpha\beta}^\gamma = N_{\alpha\bar\gamma}^{\bar\beta}$ from the general algebraic theory of anyons\cite{kitaev06}. From this fact, we derive $N_{\alpha \bar{\alpha}}^{q_i} = N_{\alpha \bar q_i}^\alpha = 0,1$, where the last equality follows from property (\ref{qa}).
As a corollary, we see that the charges $q_i$ appearing in (\ref{E1}) are exactly those that follow the fusion rule $\alpha\times q_i = \alpha$.

With the above properties, we now prove two claims. The first claim states that for any two excitations $\alpha, \alpha'$ with $\phi_{\alpha'}=\phi_\alpha$, there exists at least one
charge $q$ with $\alpha' = \alpha \times q$. The second claim is more complicated. To explain it, consider two excitations $\alpha, \beta$ in a fusion channel $\gamma$, and two other excitations $\alpha' ,\beta'$ in a fusion channel $\gamma'$. The claim states if $\phi_{\alpha'} = \phi_{\alpha}$ and $\phi_{\beta'} = \phi_\beta$ then there exist charges $q_1$ and $q_2$ such that $\alpha' = \alpha \times q_1$, $\beta' = \beta \times q_2$ and $\gamma' = \gamma \times q_1 \times q_2$.

To prove the first claim, consider an arbitrary excitation $\alpha'$ with $\phi_{\alpha'} = \phi_\alpha$. Imagine fusing together $\alpha, \bar\alpha, \alpha'$. We can do the fusion in two different orders,
\begin{equation}
(\alpha\times\bar\alpha)\times\alpha' = (\emptyset + q_1 + \dots )\times \alpha', \label{E2}
\end{equation}
and
\begin{equation}
\alpha\times(\bar\alpha\times\alpha') = \alpha\times(q_1' + q_2'+ \dots  ), \label{E3}
\end{equation}
where the ``$\dots$''  means some charges. Since the order of fusion can't affect the final result, we conclude that $\alpha'=\alpha\times q_i'$, where $q_i'$ is one of the charges in (\ref{E3}). Hence, the claim holds.

To prove the second claim, let $\alpha,\beta, \alpha', \beta'$ be any excitations with $\phi_{\alpha'} = \phi_{\alpha}$ and $\phi_{\beta'} = \phi_\beta$. Let $\gamma$
be one of the fusion channels of $\alpha, \beta$ and let $\gamma'$ be one of the fusion channels of $\alpha' ,\beta'$. Our task is to show that, given any state in $\mathbb V_{\alpha \beta}^\gamma$, we can construct at least one state in $\mathbb V_{\alpha'\beta'}^{\gamma'}$ by fusing charges onto $\alpha$ and $\beta$. To show this, we note that
$\gamma' = \gamma \times q$ for some charge $q$, by the first claim, proven above. Let us consider the fusion product $(\alpha' \times \beta') \times (\bar{\alpha} \times \bar{\beta}) \times \bar{q}$. The vacuum fusion channel must appear at least once in this fusion product since
\begin{equation}
(\alpha' \times \beta') \times (\bar{\alpha} \times \bar{\beta}) \times \bar{q} = (\gamma' + \dots) \times (\bar{\gamma} + \dots) \times \bar{q}
\end{equation}
and $\gamma' \times \bar{\gamma} \times \bar{q}$ contains the vacuum. We can now reorder the fusion product as
\begin{equation}
(\alpha' \times \bar{\alpha}) \times (\beta' \times \bar{\beta}) \times \bar{q}
\end{equation}
This reordering shouldn't change the result so the vacuum must also appear in this product. We conclude $q$ can be written as a product $q = q_1 \times q_2$ where $q_1$ appears in the fusion product of $\alpha' \times \bar{\alpha}$ and $q_2$ belongs to the fusion product of $\beta' \times \bar{\beta}$. Clearly $q_1$ and $q_2$ satisfy $q_1 \times \alpha = \alpha'$ and $q_2 \times \beta = \beta'$. Also, we know that $q_1 \times q_2 \times \gamma = q \times \gamma = \gamma'$. This proves the claim.

\section{A property of $R_{\alpha\alpha}^\beta$}

\label{sec:appd_Raa}
\begin{figure*}
\includegraphics{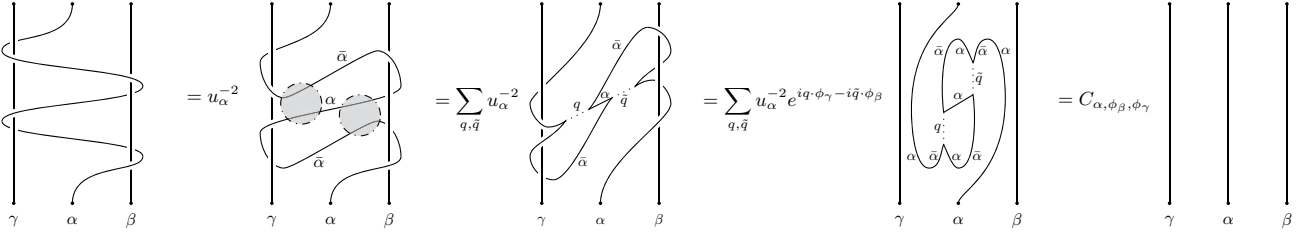}
\caption{Diagrammatic proof that the unitary matrix associated with the braiding process in the definition of $\Theta_{ijk}$ is an Abelian phase.}\label{fig_thetaijk}
\end{figure*}

In this Appendix, we show that when an excitation is braided around another identical excitation in a 2D gauge theory with group $G = \prod_{i=1}^K \mathbb Z_{N_i}$, the resulting unitary transformation is a pure phase, i.e., proportional to the identity matrix. Note that this result only holds for a full braiding: the unitary transformation associated with an exchange of two identical excitations need not be a pure phase.

Consider an arbitrary excitation $\alpha$ in a 2D gauge theory with group $G = \prod_{i=1}^K \mathbb Z_{N_i}$. In the first step, we show that the unitary transformation associated with braiding $\alpha$ around its antiparticle $\bar{\alpha}$ is a pure phase. To see this, note that according to Appendix \ref{sec:appd_fusionprop}, the only fusion outcomes for $\alpha$ and $\bar{\alpha}$ are charges: $\alpha\times\bar\alpha = \emptyset + q + \dots$. For any $q$ that appears in this fusion rule, we can use the formula (\ref{appdf_braiding}) to derive
\begin{equation}
R_{\bar\alpha\alpha}^q R_{\alpha\bar\alpha}^q  =e ^{i2\pi (s_q - s_\alpha - s_{\bar\alpha })} =  e^{-i4\pi s_\alpha},
\end{equation}
where we have used the facts ${\rm dim}({\mathbb V_{\alpha \bar \alpha}^q})=1$,  $s_q=0$ and $s_\alpha=s_{\bar\alpha}$. Examining the above identity, we can see that the statistical phase associated with braiding $\alpha$ around $\bar\alpha$ is independent of the fusion channel $q$. Hence, braiding $\alpha$ around $\bar\alpha$ gives a pure phase.

Next, we show that all the charges $q$ that appear in the fusion product $\alpha\times\bar\alpha$ have vanishing braiding statistics with $\alpha$, i.e., $\theta_{\alpha q} = 0$. To see this, imagine we have two excitations $\alpha$ and $\bar\alpha$ in the vacuum fusion channel. If we now fuse $q$ to $\bar\alpha$, the excitation $\bar{\alpha}$ will remain unchanged (i.e. $q \times \bar{\alpha} = \bar{\alpha}$) but after this fusion process, the two excitations $\alpha$ and $\bar{\alpha}$ will be in the fusion channel $q$. Let us imagine braiding $\alpha$ around $\bar\alpha$ before and after fusing $q$ into $\bar\alpha$. Clearly the two processes will
differ by a phase factor $e^{i \theta_{\alpha q}}$. Hence
\begin{equation}
R_{\bar\alpha\alpha}^qR_{\alpha\bar\alpha}^q = R_{\bar\alpha\alpha}^\emptyset R_{\alpha\bar\alpha}^\emptyset e^{i \theta_{\alpha q}}.
\end{equation}
Then since $R_{\bar\alpha\alpha}^q R_{\alpha\bar\alpha}^q$ is independent of $q$, we derive $\theta_{\alpha q} =0$.

To complete the argument, we consider a process in which $\alpha$ is braided around a pair of $\alpha$ and $\bar\alpha$ excitations. Independent of the fusion channel of the $\alpha$ and $\bar\alpha$ excitations, the unitary transformation associated with this process must be the identity since $\theta_{\alpha q} = 0$. At the same time, this braiding process can be divided into two pieces: first $\alpha$ is braided around $\bar\alpha$, and then around another $\alpha$. Since the first piece is a pure phase $e^{-i4\pi s_\alpha}$, the second must also be a pure phase. This proves the claim. In addition, we have derived the formula
\begin{equation}
R_{\alpha\alpha}^\beta R_{\alpha\alpha}^\beta = e^{i4\pi s_\alpha}
\end{equation}
where $\beta$ is any excitation in the fusion product $\alpha\times\alpha$.

\section{Proving $\Theta_{ijk}$ is well defined}
\label{sec:appd_thetaijk}

In this Appendix, we prove that $\Theta_{ijk}$ is a well defined quantities. More specifically, we show that (i) the unitary transformation associated with the braiding process defining $\Theta_{ijk}$ is always an Abelian phase even if the vortices are non-Abelian; we also show that (ii) the Abelian phase is a function of $i,j,k$ only and does not depend on the choice of vortices $\alpha, \beta, \gamma$ as long as they carry fluxes $\frac{2\pi}{N_i}e_i, \frac{2\pi}{N_j}e_j, \frac{2\pi}{N_k}e_k$ respectively.

To prove points (i) and (ii), we make use of a diagrammatic technique to compute the unitary matrix associated with the braiding process in the definition of $\Theta_{ijk}$ (for more details about this diagrammatic technique, see Ref.~\onlinecite{kitaev06}.) The technique uses space-time trajectories, where the arrow of time is drawn upward. We will not use the technique to carry out an actual calculation but only to show that the unitary transformation associated with $\Theta_{ijk}$ is a pure phase. We will make use of two diagrammatic relations in our proof. The first relation is
\begin{equation}
\includegraphics{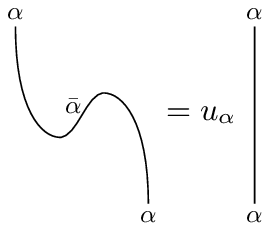}\label{diagrelation1}
\end{equation}
This relation allows us to turn downward a trajectory of $\alpha$ by introducing its antiparticle $\bar\alpha$, with a compensation of a complex factor $u_\alpha$. The factor $u_\alpha$ is related to the quantum dimension $d_\alpha$ by $d_\alpha = |u_\alpha|^{-1}$. The second relation is
\begin{equation}
\includegraphics{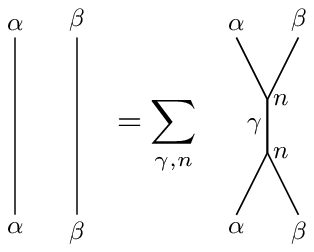}\label{diagrelation2}
\end{equation}
which means that the propagation of two particles $\alpha, \beta$ can be decomposed into a sum over their possible fusion channels, where $\gamma$ ranges over the fusion channels in $\alpha\times\beta = \sum_\gamma N_{\alpha\beta}^\gamma \gamma$ and $n$ ranges over $1,\dots, N_{\alpha\beta}^\gamma$. The vertices
\begin{equation}
\includegraphics[scale=0.8]{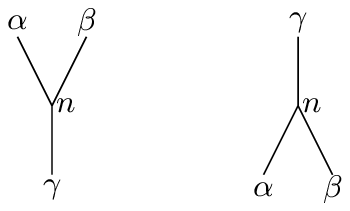}\nonumber
\end{equation}
mean the $n$th way of splitting and fusing $\alpha,\beta$ into $\gamma$ respectively.

%\begin{figure}
%\includegraphics{fig_braidingdiagram.eps}
%\caption{Two relations in the diagrammatic technique. } \label{fig_tworelations}
%\end{figure}

Now let $\alpha,\beta,\gamma$ be vortices carrying unit flux $\frac{2\pi}{N_i}e_i,\frac{2\pi}{N_j}e_j,\frac{2\pi}{N_k}e_k$ respectively. Consider the space-time trajectories of $\alpha,\beta,\gamma$ associated with the braiding process in the definition of $\Theta_{ijk}$ (Fig.~\ref{fig_thetaijk}). Making using of the relation (\ref{diagrelation1}), we establish the first equation in Fig.~\ref{fig_thetaijk}. The second equation is established by using the relation (\ref{diagrelation2}) in the two shaded regions in the second diagram in Fig.~\ref{fig_thetaijk}. The charges $q,\tilde q$ are those appearing in the fusion rule of $\alpha\times\bar\alpha$. We have used the fact that $N_{\alpha\bar\alpha}^q=1$ for any $q$, which is proven in Appendix \ref{sec:appd_fusionprop}. To establish the third equation, we notice that $q,\tilde q$ have Abelian statistics with the vortices $\beta,\gamma$. Winding $q$ around $\gamma$ in the counterclockwise direction gives rise to a phase $e^{iq\cdot\phi_\gamma}$, and winding $\tilde q$ around $\beta$ in the clockwise direction gives rise to a phase $e^{-i\tilde q\cdot\phi_\beta}$. In the fourth diagram in Fig.~\ref{fig_thetaijk}, we see that $\beta$ and $\gamma$ are decoupled from $\alpha$, while $\alpha$ still has some ``self-interaction''. However complicated this ``self-interaction'' is, it only depends on $\alpha, q,\tilde q$, but not on $\beta, \gamma$. Therefore, we can denote everything as a complex number $C_{\alpha,\phi_\beta,\phi_\gamma}$, after the summation over $q,\tilde q$ has been performed. We see that $C_{\alpha,\phi_\beta,\phi_\gamma}$ only depends on the flux of $\beta,\gamma$, but not on the choice of $\beta,\gamma$, nor on their fusion channel. Since the overall transformation must be unitary, $C_{\alpha,\phi_\beta,\phi_\gamma}$ is a pure phase.

To complete the argument, we have to show that the factor $C_{\alpha,\phi_\beta,\phi_\gamma}=e^{i\Theta_{ijk}}$ only depends on the flux of $\alpha$ but not on the choice of $\alpha$. This can be proven by fusing charges to $\alpha$. Let the outcome of the fusion be a vortex $\alpha'$. According to the Aharanov-Bohm law, one can easily see that
\begin{align}
C_{\alpha',\phi_\beta,\phi_\gamma} & =C_{\alpha,\phi_\beta,\phi_\gamma}e^{iq\cdot\phi_\beta+iq\cdot\phi_\gamma-iq\cdot\phi_\beta-iq\cdot\phi_\gamma} \nonumber\\
& =C_{\alpha,\phi_\beta,\phi_\gamma}.
\end{align}
With this, we have shown that $\Theta_{ijk}$ only depends on the flux of $\alpha,\beta,\gamma$, i.e. it only depends on $i,j,k$. Hence, $\Theta_{ijk}$ is well defined.

\section{Correspondence between labels $(a,\rho)$ and physical notions of gauge flux and gauge charge}
\label{appd:correspd}

In this appendix, we show how to translate between the mathematical labels $\alpha=(a,\rho)$, used to denote excitations in 2D Dijkgraaf-Witten models, and the physical notions of gauge flux and gauge charge. As discussed in the main text, the basic outline of correspondence is simple: the first component $a$ describes the amount of flux carried by the excitation $\alpha$, while the second component $\rho$ is related to the amount of charge attached to $\alpha$. We now explain how this works in more detail.

For each group element $a = (a_1,...,a_K)$, the corresponding gauge flux is given by $\phi = (\phi_1,...,\phi_K)$ where $\phi_i = \frac{2\pi}{N_i} a_i$. Likewise, for each representation $\rho$ we should define a corresponding gauge charge $q = (q_1,...,q_K)$. This correspondence is easy to define for the case where $\alpha$ is a pure charge
excitation: $\alpha = (0,\rho)$. Indeed in this case, Eq. (\ref{projective}) implies that $\rho$ is a \emph{linear} representation of $G$ --- provided that we choose a ``gauge''\cite{footnote5} such that $\omega(a,b,c) = 1$ if any of $a,b,c$ is $0$. It follows that $\rho$ can be written in the form $\rho(h) = \exp(\sum_k \frac{2\pi i}{N_k} q_k h_k)$ for some integer vector $(q_1,...,q_K)$. This defines the desired correspondence $\rho \leftrightarrow q = (q_1,...,q_K)$.

How does the correspondence work for vortex excitations $\alpha = (a,\rho)$ where $a \neq 0$? In this case, $\rho$ is a \emph{projective} representation, so there is no natural way to translate $\rho$ into an integer vector $(q_1,...,q_K)$. This is related to the general point made in Section \ref{sec:2dabelian1}: we do not know a physically meaningful way to define the absolute charge carried by a vortex excitation. On the other hand, if we \emph{compare} two vortex excitations with the same flux, $\alpha = (a,\rho)$ and $\alpha' = (a,\rho')$, and we find that $\rho,\rho'$ are related by $\rho'(h) = \rho(h) \cdot \exp(\sum_k \frac{2\pi i}{N_k} q_k h_k)$ for some $(q_1,...,q_k)$, then we can say that $\alpha'$ can be obtained from $\alpha$ by attaching charge $q=(q_1,...,q_K)$.

\section{Showing that all solutions to the constraints (\ref{3dconstraints}) can be written in the form (\ref{theta_value_3d1}), (\ref{theta_value_3d2})}
\label{sec:appd_formula}

In this section, we show that if $\Theta_{i,l}$ and $\Theta_{ij,l}$ obey the constraints (\ref{3dconstraint_exch})-(\ref{3dconstraint_cyclic3}), then they can be written in the form (\ref{theta_value_3d1})-(\ref{theta_value_3d2}), i.e.,
\begin{align}
\Theta_{i,l} &= \frac{2\pi}{N_{il}}(M_{ili}-M_{lii}) \label{theta_value_3d_app} \\
\Theta_{ij,l} &= \frac{2\pi N^{ij}}{N_{il} N_j}(M_{ilj}- M_{lij}) + \frac{2\pi N^{ij}}{N_{jl} N_i}(M_{jli}- M_{lji}) \nonumber
\end{align}
for some integer tensor $M_{ijk}$.
The first step is rewrite the second equation as
\begin{align}
\Theta_{ij,l} = \frac{2\pi N^{ijl}}{N^{jl} N_{ijl}}(M_{ilj}- M_{lij}) + \frac{2\pi N^{ijl}}{N^{il} N_{ijl}}(M_{jli}- M_{lji})
\label{theta_value_3d_app_2}
\end{align}
This new form of the second equation can be derived from the relations
\begin{align}
\frac{N^{ij}}{N_{il} N_j} =  \frac{N^{ijl}}{N^{jl} N_{ijl}}, \quad \frac{N^{ij}}{N_{jl} N_i} =\frac{N^{ijl}}{N^{il} N_{ijl}}
\end{align}
which in turn follow from the identities
\begin{align}
N^{ij} = \frac{N_i N_j}{N_{ij}}, \quad N^{ijl}= \frac{N^{ij} N^{il} N^{jl} N_{ijl}}{N_i N_j N_l}
\end{align}

We now construct the integer tensor $M_{ijk}$. First, we set $M_{iii} = 0$ for all $i$, and we set
\begin{align}
M_{lii} = M_{iil} = 0, \quad M_{ili} = \frac{N_{il} }{2\pi} \Theta_{i,l},
\end{align}
for all $i \neq l$. Next, for each $i < j < l$, we define
\begin{align}
M_{ilj} &= - b \frac{N_{ijl}}{2\pi} \Theta_{ijl} , \nonumber \\
M_{jil} &=  \frac{N_{ijl}}{2\pi} (a \Theta_{jli} + b \Theta_{lij}), \nonumber \\
M_{lji} &= - a \frac{N_{ijl}}{2\pi} \Theta_{ijl}, \nonumber \\
M_{lij} &= 0, \quad M_{jli} = 0, \quad M_{ijl} = 0
\end{align}
where $a$ and $b$ are integers such that
\begin{equation}
a \frac{N^{ijl}}{N^{il}} - b \frac{N^{ijl}}{N^{jl}} = 1
\label{abrel}
\end{equation}
(The existence of $a,b$ will be established below). If one substitutes the tensor $M_{ijk}$ into (\ref{theta_value_3d_app}, \ref{theta_value_3d_app_2}), it is straightforward to check that the resulting expressions exactly reproduce the invariants $\Theta_{i,l}$ and $\Theta_{ij,l}$ as long as these invariants obey the
constraints (\ref{3dconstraints}).

At this point, we have successfully constructed a tensor $M_{ijk}$ that satisfies equations (\ref{theta_value_3d_app}, \ref{theta_value_3d_app_2}). However, there are two gaps in our derivation that need to be addressed. First, we have to show that the components of $M_{ijk}$ are all integers. Second, we have to show that we can always find integers $a,b$ satisfying (\ref{abrel}). The fact that the components of $M_{ijk}$ are all integers is easy to prove, as it follows immediately from the two constraints (\ref{3dconstraint_quant_ex}), (\ref{3dconstraint_quant}). As for the second statement, this will follow if we can show that $N^{ijl}/N^{il}$ and $N^{ijl}/N^{jl}$ are relatively prime. The latter property can be derived from a simple observation: we note that the only prime factors appearing in $N^{ijl}/N^{il}$ are
those that divide into $j$ more times than either $i$ or $l$. Similarly, the only prime factors appearing in $N^{ijl}/N^{jl}$ are those that
divide into $i$ more times than $j$ or $l$. We conclude that these two numbers do not share any prime factors so they are relatively prime.

\section{Counting the number of values of $\Theta_{i,l}, \Theta_{ij,l}, \Theta_{ijk,l}$}
\label{sec:appd_counting}
In this section, we consider the formulas  (\ref{theta_value_3d1})-(\ref{theta_value_3d3}), reprinted below for convenience:
\begin{subequations}
\begin{align}
\Theta_{i,l} &= \frac{2\pi}{N_{il}}(M_{ili}-M_{lii}) \label{theta_value_3d1_app}\\
\Theta_{ij,l} &= \frac{2\pi N^{ijl}}{N^{jl} N_{ijl}}(M_{ilj}- M_{lij}) +\frac{2\pi N^{ijl}}{N_{il} N_{ijl}}(M_{jli}- M_{lji})\label{theta_value_3d2_app} \\
\Theta_{ijk,l} & =\frac{2\pi}{N_{ijkl}} \sum_{\hat p} {\rm sgn}(\hat p) L_{\hat p(i)\hat p(j)\hat p(k)\hat p(l)} \label{theta_value_3d3_app}
\end{align}
\end{subequations}
What we will show is that invariants $\Theta_{i,l}, \Theta_{ij,l}, \Theta_{ijk,l}$ take on at least
\begin{equation}
\prod_{i<j} \left(N_{ij}\right)^2 \prod_{i<j<l} \left(N_{ijl}\right)^2 \prod_{i<j<k<l} N_{ijkl}. \label{appd_e1}
\end{equation}
different values when $M_{ijk}$ ranges over all integer tensors, and $L_{ijkl}$ ranges over all integer tensors obeying $L_{ijkl} =0$ if $i,j,k$ are not all distinct.
(The reader may notice that the second equation (\ref{theta_value_3d2_app}) differs slightly from equation (\ref{theta_value_3d2}) in the main text. The equivalence between these two equations follows from simple identities and is explained in appendix \ref{sec:appd_formula}).

To perform our counting, we consider separately each of the components of $\Theta_{i,l}, \Theta_{ij,l}$. First, we consider the invariant $\Theta_{i,l}$ for $i \neq l$. From equation (\ref{theta_value_3d1_app}) we can see that $\Theta_{i,l}$ can take on $N_{il}$ different values as we vary $M_{ili}, M_{lii}$. We can also see that the different $\Theta_{i,l}$ invariants with $i \neq l$ are independent of one another. Hence, all together the $\Theta_{i,l}$ invariants can take on at least
\begin{equation}
\mathcal N_1 = \prod_{i \neq l} N_{il}  = \prod_{i < l} \left(N_{il}\right)^2
\end{equation}
different values. Next, we fix $i < j < l$ and consider the three invariants $\Theta_{ij,l}, \Theta_{jl,i}, \Theta_{lij}$. We will show that these invariants can take on at least $(N_{ijl})^2$ different values. To see this, we set
\begin{align}
M_{ilj} &= -b x, \quad M_{lji} = -a x , \quad M_{jil} = y \nonumber \\
M_{lij} &= 0, \quad M_{jli} = 0, \quad M_{ijl} = 0
\end{align}
where $x,y$ are arbitrary integers and $a$ and $b$ are chosen so that
\begin{equation*}
a \frac{N^{ijl}}{N^{il}} - b \frac{N^{ijl}}{N^{jl}} = 1
\end{equation*}
(One can always find such $a,b$ since $\frac{N^{ijl}}{N^{il}}$ and $\frac{N^{ijl}}{N^{jl}}$ are relatively prime, as explained at the end of appendix \ref{sec:appd_formula}). Substituting these expressions for $M$ into (\ref{theta_value_3d2_app}), it is straightforward to derive the following two formulas:
\begin{eqnarray}
\Theta_{ij,l} &=& 2\pi x/N_{ij,l}  \nonumber \\
a \Theta_{jl,i} + b \Theta_{li,j} &=& 2\pi y/N_{ijl}
\end{eqnarray}
From these formulas it is clear that $\Theta_{ij,l}, \Theta_{jl,i}, \Theta_{li,j}$ take on different values for each $x,y$ with $0 \leq x,y \leq N_{ijl} - 1$. Hence, $\Theta_{ij,l}, \Theta_{jl,i}, \Theta_{li,j}$ can take on at least $N_{ijl}^2$ different values, as claimed above. It is also clear that the values for different $i,j,l$
are independent of one another. Hence, all together the $\Theta_{ij,l}$ invariants can take on at least
\begin{equation}
\mathcal N_2 = \prod_{i<j<l} \left(N_{ijl}\right)^2,
\end{equation}
different values.
Finally, we consider the invariants $\Theta_{ijk,l}$ for $i < j < k < l$. From equation (\ref{theta_value_3d3_app}), it is clear that $\Theta_{ijk,l}$ can take on
$N_{ijkl}$ different values as we vary $L_{ijkl}$. Hence, all together, the number of distinct values of $\{\Theta_{ijk,l}\}$ is at least
\begin{equation}
\mathcal N_3 = \prod_{i<j<k<l} N_{ijkl}.
\end{equation}
Combining all the cases, we conclude that the total number of values that the invariants can take on is at least $\mathcal N_1\mathcal N_2\mathcal N_3$, which is what we wanted to show. (In fact, with more careful accounting, one can show that this inequality is actually an equality, but we will not need this sharper result here).


\begin{thebibliography}{100}
\bibitem{footnote1} We explain the logic behind this claim in Sec.~\ref{sec:model_phase}.
\bibitem{footnote2} Besides simple loops such as circles or linked circles, loops can be knotted or even branched. In this paper, we only consider simple loops without knotting or branching.
\bibitem{footnote4} Note that this correspondence between loop and particle braiding only holds if the loops are linked to a base loop $\gamma$; the braid group for unlinked loops is different from the 2D case, as discussed in Ref.~\onlinecite{baez07}.
\bibitem{footnote5} It is not hard to see that for any cohomology class in $H^3(G,U(1)$, we can always find a representative cocycle $\omega$ with the property that $\omega(a,b,c) = 1$ if any of $a,b,c$ is $0$.
\bibitem{footnote6} It can be shown that the Dijkgraaf-Witten model corresponding to the cocycle $\omega$ (\ref{omega3d}) is Abelian if $L_{ijkl} = 0$.
\bibitem{footnote10}Here, when we say braiding statistics, we mean the complete set of algebraic data for anyon systems, including quantum dimensions and fusion rules. For more details, see e.g. Appendix E of Ref. \onlinecite{kitaev06}.
\end{thebibliography}
\end{document}